\documentclass[12pt]{article}
\usepackage{tocloft}
\usepackage{amsfonts}
\usepackage{amsmath}
\usepackage{amssymb}
\usepackage{array}
\usepackage{bigints}
\usepackage{bm}
\usepackage{booktabs}
\usepackage[nosort]{cite}
\usepackage{color}
\usepackage{dsfont}
\usepackage{float}
\usepackage{framed}
\usepackage{graphicx}
\usepackage{indentfirst}
\usepackage{mathrsfs}
\usepackage{multirow}
\usepackage{pdflscape}
\usepackage{setspace}
\usepackage{subdepth}
\usepackage{subfig}
\usepackage{titlesec}
\usepackage{wrapfig}
\usepackage[all]{xy}
\usepackage{young}
\usepackage[vcentermath]{youngtab}
\usepackage{relsize}
\usepackage{stackengine}
\usepackage{verbatim}
\usepackage{slashed}
\usepackage[utf8]{inputenc}

\usepackage{hyperref}
\hypersetup{colorlinks=true}
\hypersetup{linkcolor=black}
\hypersetup{citecolor=black}
\hypersetup{urlcolor=black}

\numberwithin{equation}{section}

\usepackage[left=2.5cm,right=2.5cm,top=2.5cm,bottom=3cm]{geometry}
\fontdimen2\font=1.2\fontdimen2{\jot}{5pt}

\titleformat{\section}{\large\bfseries}{\thesection.}{4pt}{}
\titlespacing{\section}{0pt}{20pt}{6pt}

\titleformat{\subsection}{\normalfont\bfseries}{\thesubsection.}{4pt}{}
\titlespacing{\subsection}{0pt}{15pt}{6pt}

\titleformat{\subsubsection}{\normalfont\itshape}{\thesubsubsection.}{4pt}{}
\titlespacing{\subsubsection}{0pt}{15pt}{6pt}

\titleformat{\paragraph}{\normalfont\itshape}{\theparagraph.}{4pt}{}
\titlespacing{\paragraph}{0pt}{15pt}{6pt}

\newcommand{\bea}{\begin{eqnarray}}
\newcommand{\eea}{\end{eqnarray}}
\newcommand{\beq}{\begin{equation}}
\newcommand{\eeq}{\end{equation}}

\def\<{\langle}
\def\>{\rangle}

\def\nn{\nonumber}

\def\cG {{\cal G}}

\def\cS {{\cal S}}

\def\cS {{\cal S}}
\def\SL {{SL(2,\mathbb{R})}}

\def\cO {{\mathcal O}}
\def\cT {{\mathcal T}}

\DeclareFontShape{OT1}{cmr}{mx}{n}%
{<->cmr10}{}
\newcommand{\mytitlefont}{\fontseries{mx}\selectfont}
\DeclareMathAlphabet{\titlemath}{OT1}{cmr}{mx}{n}

\begin{document}

\begin{titlepage}

\begin{center}
			
~\\[0.7cm]
			
{\fontsize{24pt}{0pt} \mytitlefont Large $N$ analytical functional bootstrap I: 1D CFTs and total positivity}
			
~\\[0.4cm]

Zhijin Li $^{1,2}$

~\\[0.1cm]

$^1$\,{\it Shing-Tung Yau Center and School of Physics, Southeast University , Nanjing 210096, China}~\\[0.2cm]

$^2$\,{\it Department of Physics, Yale University, New Haven, CT 06511, USA}\\[0.2cm]

\end{center}

\vskip0.5cm
			
\noindent We initiate the analytical functional bootstrap study of conformal field theories with large $N$ limits. In this first paper we particularly focus on the 1D 
$O(N)$ vector bootstrap. 
We obtain a remarkably simple  bootstrap equation from the $O(N)$ vector crossing equations in the large $N$ limit. The bootstrap bound is saturated by the generalized free field theory.
We study the analytical extremal functionals of this crossing equation, for which the total positivity of the $SL(2,\mathbb{R})$ conformal block plays a critical role. We prove the 
$SL(2,\mathbb{R})$ conformal block is totally positive for large scaling dimension $\Delta$ and  show that the total positivity is violated below a critical value $\Delta_{\textrm{TP}}^*\approx 0.32315626$. 
The conformal block forms a surprisingly sophisticated  mathematical structure, which for instance can violate total positivity at the order $10^{-5654}$ for a normal value $\Delta=0.1627$!
We construct a series of analytical functionals $\{\alpha_M\}$ which 
satisfy the bootstrap positive conditions up to a range $\Delta\leqslant \Lambda_M$. 
The functionals $\{\alpha_M\}$ have a trivial large $M$ limit. Surprisingly, due to total positivity, they can approach the large $M$ limit in a way consistent with the 
bootstrap positive conditions for arbitrarily high $\Lambda_M$, therefore proving the bootstrap bound analytically. 
Our result provides a concrete example to illustrate how the  analytical properties of the conformal block lead to nontrivial bootstrap bounds. We expect this work paves the way for large $N$ analytical functional bootstrap in higher dimensions.


\end{titlepage}

	
\setcounter{tocdepth}{3}
\renewcommand{\cfttoctitlefont}{\large\bfseries}
\renewcommand{\cftsecaftersnum}{.}
\renewcommand{\cftsubsecaftersnum}{.}
\renewcommand{\cftsubsubsecaftersnum}{.}
\renewcommand{\cftdotsep}{6}
\renewcommand\contentsname{\centerline{Contents}}
	
\tableofcontents

\newpage
\section{Introduction}
The conformal bootstrap \cite{Ferrara:1973yt, Polyakov:1974gs}
has been revived since the breakthrough work \cite{Rattazzi:2008pe}, which shows that strong constraints on the parameter space of general conformal field theories (CFTs) can be obtained merely from few consistency conditions. This approach has led to remarkable successes in studying the strongly coupled critical phenomena, see \cite{Poland:2018epd, Poland:2022qrs} for comprehensive reviews. It is followed by an immediate  question: {\it how can such strong results be obtained from such few inputs?} The bootstrap results, or the ``numerical experiments" indicate certain mysterious mathematical structures in conformal theories which can play key roles in determining the CFT landscape. Since the ingredients in conformal bootstrap are just  unitarity and the conformal blocks of the conformal group $SO(D+1,1)$, the unreasonable effectiveness of the bootstrap method is likely related to certain properties of the $SO(D+1,1)$ conformal blocks.
The goal of this work is to explore such presumed mathematical structures. We will focus on 1D large $N$ conformal bootstrap for which the  extremal bootstrap functional can be studied analytically. Moreover, we will clarify the key mathematical property which makes our construction possible. 

We will start with the 1D $O(N)$ vector bootstrap in the large $N$ limit. Under suitable conditions the $O(N)$ vector crossing equations are reduced to one of the simplest bootstrap equations
$$ 
\sum_{\cO\in S}\lambda_{\cO}^2 \,z^{-2\Delta_\phi}\,G_{\Delta}(z) - \sum_{\cO \in T}\lambda_{\cO}^2 \,(1-z)^{-2\Delta_\phi}\,G_{\Delta}(1-z)=0,   $$
where $G_\Delta(z)$ is the 1D $\SL$ conformal block.
The bootstrap bound on the scaling dimension of the lowest operator in the $O(N)$ traceless symmetric sector ($T$) is saturated by the generalized free field theory. 
We find the key mathematical property responsible for the bootstrap constraints can be provided by the {\it total positivity} of the $\SL$ conformal block:
\vspace{1mm}
\begin{center}
\fbox{ 
    Total positivity of the $\SL$ conformal block $\longrightarrow$ Large $N$ bootstrap bound.    
}\end{center}
\vspace{1mm}
We will show that total positivity of the $\SL$ conformal block relates to a surprisingly delicate mathematical structure.
Based on the total positivity of the $\SL$ conformal block, we construct a series of analytical functionals for above bootstrap equation which can satisfy the bootstrap positive conditions up to arbitrarily high scaling dimension. 
 
Our interests in the large $N$ CFTs and their bootstrap studies are motivated by several reasons.

The large $N$ CFTs  play fundamental roles in the AdS/CFT correspondence \cite{Maldacena:1997re, Witten:1998qj,Gubser:1998bc}. In the large $N$ limit, the conformal correlation functions are dominated by the generalized free field theories, and they provide pivotal solutions to the conformal crossing equations \cite{Fitzpatrick:2012yx, Komargodski:2012ek, Pal:2022vqc}. Perturbative CFT data can be obtained by expanding the solutions to the crossing equations near generalized free field theories \cite{Gopakumar:2016wkt, Alday:2016njk, Penedones:2019tng}. The role of large $N$ CFTs in holography has been extensively studied, e.g.  \cite{Heemskerk:2009pn, Fitzpatrick:2011dm}. The generalized free field theories also provide nice examples for the harmonic analysis of the Euclidean conformal group \cite{Karateev:2018oml}. In this work, we will show that the generalized free field theories are not just pivotal solutions to the crossing equations, but can also saturate the bootstrap bounds. This indicates a special positive structure in the generalized free field theories which restricts any dynamical corrections to the bounded parameters are either vanishing or negative. Decoding the positive structure is an interesting problem for bootstrap studies.

We use $O(N)$ vector bootstrap to study the large $N$ CFTs. The $O(N)$ vector bootstrap plays a special role in conformal bootstrap with global symmetries. Due to novel algebraic relations between crossing equations with different global symmetries   \cite{Li:2020bnb, Li:2020tsm}, the non-$O(N)$ vector crossing equations can be mapped to those of $O(N)$ vector's, and their bootstrap bounds are identical or weaker than the $O(N)$ vector bootstrap bounds.
On the physics side, the $O(N)$ vector bootstrap bounds have close relation to several interesting theories. For instance,  the 3D $O(N)$ vector bootstrap bounds have two types of kinks \cite{Kos:2013tga,Li:2018lyb}. The type I kinks with an $O(N)$ vector scalar $\phi$ near a free boson $\Delta_\phi=\frac{1}{2}$ are related to the critical $O(N)$ vector model \cite{Kos:2013tga}, while the type II kinks with $\Delta_\phi$ near free fermion bilinears also appear in general dimensions and show close relation with conformal gauge theories \cite{Li:2020bnb,Li:2018lyb}.

The analytical construction of the extremal bootstrap  functional \cite{El-Showk:2012vjm} provides a substantial approach to uncover the positive structure in conformal bootstrap.
Analytical extremal functionals have been firstly constructed in \cite{Mazac:2016qev, Mazac:2018mdx, Mazac:2018ycv} for a 1D conformal bootstrap problem, in which the bootstrap bound is saturated by the generalized free fermion theory \cite{Gaiotto:2013nva}. Analytical functionals for higher dimensional conformal bootstrap have been studied in 
\cite{Paulos:2019gtx, Mazac:2019shk, Caron-Huot:2020adz}. In \cite{Mazac:2019shk} a family of functional basis dual to the generalized free field spectrum has been constructed, which shows close relation to the conformal dispersion relation \cite{Carmi:2019cub}. Nevertheless, it remains a puzzle to realize the positive conditions, which are the crucial ingredients for conformal bootstrap. In this work, we aim to answer this critical question for the large $N$ analytical functional bootstrap in a simplified laboratory, the 1D $O(N)$ vector bootstrap. In 1D CFTs, there is only one conformal invariant cross ratio $(z)$ and the spectrum does not depend on spin. 
Interestingly, although the crossing equations have been simplified notably in 1D, the bootstrap bounds show similar patterns as their higher dimension analogs. Therefore we expect the 1D analytical functional bootstrap studies are instructive for similar studies in higher dimensions.

In addition, the 1D conformal bootstrap with global symmetries also corresponds to many interesting physics problems. A large set of 1D CFTs are given by the line defects of higher dimensional CFTs. Two typical examples are provided by the monodromy line defect in the 3D Ising model \cite{Billo:2013jda, Gaiotto:2013nva} and the Wilson lines in the 4D $N=4$ SYM \cite{Giombi:2018qox,Liendo:2018ukf, Cavaglia:2021bnz, Ferrero:2021bsb,Cavaglia:2022qpg}. The 1D CFTs can also be realized as boundary theories of quantum field theories in AdS$_2$ background \cite{Paulos:2016fap,Antunes:2021abs}.
Recently, there are growing interests in the applications of 1D $O(N)$ symmetric CFTs in the celestial holography \cite{Garcia-Sepulveda:2022lga, Jiang:2022hho}. Conformal bootstrap in 1D has been  a powerful approach to extract dynamical information in above theories.

Total positivity of the 1D $\SL$ conformal block will play a key role in constructing the analytical functionals of the 1D large $N$ bootstrap. In mathematics the total positivity has been extensively studied since the early of 20th century. It has deep connections to quantum field theories, see e.g. \cite{Arkani-Hamed:2012zlh,Arkani-Hamed:2013jha,Arkani-Hamed:2020blm,Herrmann:2022nkh}. The possible role of total positivity in conformal bootstrap has been proposed in \cite{Arkani-Hamed:2018ign}, in which the authors focused on the geometrical configuration supported by the $\SL$ conformal blocks. Due to total positivity,  the 1D bootstrap equation (without an $O(N)$ global symmetry) admits a cyclic polytope structure which can lead to nontrivial constraints on the CFT data, see also \cite{Sen:2019lec, Huang:2019xzm}. In this work, we will focus on a new 1D bootstrap equation with a different approach, but we will reach a similar conclusion that the total positivity of the $\SL$ conformal block can play a key role for the bootstrap constraints.

This paper is organized as follows. In Section \ref{sec2} we study the 1D $O(N)$ vector numerical bootstrap with large $N$. We discuss similarities and differences between the 1D and higher dimensional $O(N)$ vector bootstrap.  
We obtain a simplified crossing equation, which determines the first part of the $O(\infty)$ vector bootstrap bound and provides an ideal example for analytical functional bootstrap study.
In Section \ref{sec3} we study total positivity of the $\SL$ conformal block, which will be important to construct the analytical functionals. 
In Section \ref{sec4} we construct the analytical functionals for the 1D $O(\infty)$ vector bootstrap which is saturated by the generalized free field theory. We firstly review the functional basis for 1D conformal block obtained from the dispersion relation. Then we explain how the total positivity of the conformal block function can play a key role to construct the analytical functionals satisfying the bootstrap positivity conditions. This work initiates a series of analytical functional bootstrap studies of the large $N$ CFTs and their holographic duals, for which we briefly discuss in Section \ref{sec5}.

\section{Large $N$ numerical conformal bootstrap in 1D}\label{sec2}
In this section we study  1D $O(N)$ vector numerical conformal bootstrap in the large $N$ limit.  
The 1D  conformal bootstrap has been studied in \cite{Gaiotto:2013nva, Cavaglia:2021bnz, Antunes:2021abs,Cavaglia:2022qpg,Paulos:2019fkw, Ferrero:2019luz,Ghosh:2021ruh,Gimenez-Grau:2022czc}.  
Our interest in the 1D $O(N)$ vector bootstrap is from the observation that the 1D bootstrap bounds share several key properties of the $O(N)$ vector bootstrap bounds in higher dimensions, thus it can provide a drastically simplified while still representative example to study the underlying mathematical structures in conformal bootstrap.
The numerical bootstrap results  provide insightful bases for analytical functional bootstrap study in Section \ref{sec4}.

\subsection{$O(N)$ vector crossing 
 equations in 1D}
Let us consider an operator $\phi_i$ which forms a vector representation of the $O(N)$ global symmetry. Its four point correlation function is given by
\beq
\langle \phi_i(x_1)\phi_j(x_2)\phi_k(x_3)\phi_l(x_4) \rangle =\frac{1}{\left(x_{12}^2x_{34}^2\right)^{\Delta_{\phi}}}\cG_{ijkl}(z), \label{corrf}
\eeq
where the variables $x_i$ are the 1D 
coordinates, $x_{ij}=x_i-x_j$ and the conformal invariant cross-ratio $z$ is defined as
\beq
z=\frac{x_{12}x_{34}}{x_{13}x_{24}}.
\eeq
When the external operators $\phi_i(x_i)$ are in the ordered configuration $x_1<x_2<x_3<x_4$, the cross-ratio stays in the range $z\in (0,1)$. The stripped correlation function $\cG_{ijkl}(z)$ in (\ref{corrf}) can be  analytically continued in  the complex plane except the branch points at $z=0,1,\infty$, which correspond to coincidences of two operators. The $\cG_{ijkl}(z)$ is a holomorphic function with two branch cuts at $(-\infty,0]$  and $[1,+\infty)$.
In the s-channel (12)(34) limit with $z\rightarrow 0$,
the conformal correlation function $\cG_{ijkl}(z)$ can be expanded in terms of the four point invariant tensors of $O(N)$ singlet ($S$), traceless symmetric ($T$) and anti-symmetric ($A$) representations 
\beq
\cG_{ijkl}(z)=\delta_{ij}\delta_{kl}\cG^S(z)+ \left(\delta_{ik}\delta_{jl}+\delta_{il}\delta_{jk}-\frac{2}{N}\delta_{ij}\delta_{kl}\right)\cG^T(z)+ (\delta_{il}\delta_{jk}-\delta_{ik}\delta_{jl})\cG^A(z), \label{ONep}
\eeq
in which $\cG^R$ denotes the series expansion
\beq
\cG^R(z)=\sum_{\cO_R} ~ \lambda_{\cO_R}^2 G_\Delta(z)
\eeq
of the s-channel $\SL$ conformal block \cite{Dolan:2011dv}
\beq
G_\Delta(z)=z^\Delta \, _2F_1(\Delta,\Delta,2\Delta;z). \label{1DCB}
\eeq
Alternatively, one can expand the same correlation functions (\ref{ONep}) in the t-channel (23)(41) limit, which can be formally written as $\cG_{jkli}(1-z)$.
The crossing symmetry of the correlation function (\ref{corrf}) identifies the s- and t-channel expansions  
\beq
z^{-2\Delta_\phi}\cG_{ijkl}(z)=
(1-z)^{-2\Delta_\phi}\cG_{jkli}(1-z).
\eeq
Together with (\ref{ONep}), above crossing equation
leads to following independent equations 
\begin{align}
    z^{-2\Delta_\phi}\left(\cG^T(z)-\cG^A(z)\right) =&(1-z)^{-2\Delta_\phi}\left(\cG^T(1-z)-\cG^A(1-z)\right), \label{fdcr0}\\
z^{-2\Delta_\phi}\left(\cG^S(z)-\frac{2}{N}\cG^T(z)\right) =&(1-z)^{-2\Delta_\phi}\left(\cG^T(1-z)+\cG^A(1-z)\right). \label{fdcr}
\end{align}
Note to derive above crossing equations, we do not assume the statistical property of the external operator $\phi_i$, so they can be applied to both fermions and bosons in 1D.

A family of unitary solutions to the $O(N)$ vector crossing equations are provided by  
\begin{align}
    \cG^S(z) &= 1+\frac{2}{N}\,\cG^T(z), \label{GFF1}\\
\cG^T(z) &= \frac{1}{2}z^{2\Delta_\phi}\left((1-z)^{-2 \Delta_\phi}-\lambda \right), \label{GFF2}\\
\cG^A(z) &= \frac{1}{2}z^{2\Delta_\phi}\left((1-z)^{-2 \Delta_\phi}+\lambda \right), \label{GFF3}
\end{align}
in which $\lambda=\pm1$ give the $O(N)$ symmetric generalized free fermion and boson theories. Above correlation functions can be decomposed into the 1D conformal blocks
\beq
\cG^R(z)=\sum_{n=0}^\infty c_n^R ~G_{2\Delta_\phi+n}(z), \label{GFF4}
\eeq 
where
\bea
c_n^T=\frac{(2\Delta_\phi)_n^2}{2\,n!(4\Delta_\phi+n-1)_n}(1-(-1)^n\lambda), \label{OPE1}\\
c_n^A=\frac{(2\Delta_\phi)_n^2}{2\, n!(4\Delta_\phi+n-1)_n}(1+(-1)^n\lambda). \label{OPE2}
\eea
For different $\lambda$'s ($|\lambda|<1$),
the correlation functions $\cG^{S/T/A}$ contain the same spectrum $\Delta_n=2\Delta_\phi+n, ~ n\in\mathbb{N}$. An interesting question in CFT studies is that given the whole spectrum of a CFT, can we  determine the theory uniquely? The correlation function (\ref{GFF1}-\ref{GFF3})
provides a counter example for this question.

The $O(N)$ vector crossing equations (\ref{fdcr0},\ref{fdcr}) have the same algebraic structure as those in higher dimensions \cite{Rattazzi:2010yc, Kos:2013tga}. However, in higher dimensions, there are spin selection rules in different $O(N)$ representations due to the boson symmetry of the external scalars. The correlation function is invariant under permutation $(i,x_1)\leftrightarrow (j,x_2)$, which leads to
\beq
\lambda_{\phi\phi\cO_S}=c_S(-1)^\ell\lambda_{\phi\phi\cO_S}, ~~
\lambda_{\phi\phi\cO_T}=c_T(-1)^\ell\lambda_{\phi\phi\cO_T}, ~~
\lambda_{\phi\phi\cO_A}=c_A(-1)^\ell\lambda_{\phi\phi\cO_A}, \label{OPEtrs1}
\eeq
where $c_S=c_T=1,~ c_A=-1$ are the signs from $O(N)$ indices when permuting two $\phi$'s.
Therefore only even (odd) spins can appear in the $S/T$ ($A$) representations. While there is no spin in 1D, do we have similar selection rules in different $O(N)$ representations? The answer is yes and it relates to the so-called $\mathcal{S}$-parity symmetry \cite{Billo:2013jda, Gaiotto:2013nva}.

The action of the $\cS$-parity is
\beq
\cS: ~~~ x\rightarrow -x, ~~~\cS\, \cO(x)\,\cS=(-1)^{S_\cO}\cO(-x).
\eeq 
In 1D, the continuous part of the conformal symmetry preserves the cyclic order of the three point function $\langle\cO_1(x_1)\cO_2(x_2)\cO_3(x_3)\rangle $ with $ x_1<x_2<x_3$.
However, the cyclic order can be modified by the $\cS$ transformation
\beq
\cS:~~ \langle\cO_1(x_1)\cO_2(x_2)\cO_3(x_3)\rangle \rightarrow  
(-1)^{S_{\cO_1}+S_{\cO_2}+S_{\cO_3}}\langle\cO_3(-x_3)\cO_2(-x_2)\cO_1(-x_1)\rangle.
\eeq 
Therefore in an $\cS$-parity invariant theory, we have
\beq
\lambda_{\cO_1\cO_2\cO_3}=(-1)^{S_{\cO_1}+S_{\cO_2}+S_{\cO_3}}\lambda_{\cO_2\cO_1\cO_3}.
\eeq
In the 1D $O(N)$ vector bootstrap, if the external operators $\phi_i$ are scalars, the boson symmetry between the two $\phi$'s requires 
\beq
\lambda_{\phi\phi\cO_S}=(-1)^{S_{\cO_S}}\lambda_{\phi\phi\cO_S}, ~~
\lambda_{\phi\phi\cO_T}=(-1)^{S_{\cO_T}}\lambda_{\phi\phi\cO_T}, ~~
\lambda_{\phi\phi\cO_A}=-(-1)^{S_{\cO_A}}\lambda_{\phi\phi\cO_A}, \label{OPEtrs2}
\eeq
which leads to 
\beq
S_{\cO_S}=1, ~~S_{\cO_T}=1, ~~S_{\cO_A}=-1. \label{bosonP}
\eeq 
While for the external fermions, the $\cS$-parity charges in the $O(N)$ representations are opposite
\beq
S_{\cO_S}=-1, ~~S_{\cO_T}=-1, ~~S_{\cO_A}=1. \label{fermionP}
\eeq 
In the generalized free boson theory,  the $\cS$-parity of the double-trace operators $\cO_n=\phi\partial^n\phi$ is $S_n=(-1)^n$. According to the $\cS$-parity charges in (\ref{bosonP}), the generalized free boson theory has spectrum $\cO_{2n}$ with $\Delta=2\Delta_\phi+2n$ in the $S/T$  sectors and spectrum $\cO_{2n+1}$  with $\Delta=2\Delta_\phi+2n+1$ in the $A$ sector.  The spectra in $S/T$ and $A$ sectors are switched in the generalized free fermion theory due to the $\cS$-parity charges (\ref{fermionP}).

The $O(N)$ vector bootstrap plays a special role in bootstrapping CFTs with general global symmetries. In \cite{Li:2020bnb,Li:2020tsm} it has been verified that for a large variety of symmetries $\mathbf{G}$ and representations $\mathbf{R}$, the crossing equations of the four point correlator $\langle \mathbf{R}\bar{\mathbf{R}}\mathbf{R}\bar{\mathbf{R}}\rangle$ and  $\langle \mathbf{R}\mathbf{R}\bar{\mathbf{R}}\bar{\mathbf{R}} \rangle$  can be linearly mapped into the $O(N)$ symmetric form (\ref{fdcr0},\ref{fdcr}) through a transformation $\cT_\mathbf{R}$ which is consistent with positivity conditions in the bootstrap algorithm. The transformation $\cT_\mathbf{R}$ is purely algebraic so can also be applied in 1D conformal bootstrap. In consequence, the bootstrap bound on the lowest $\mathbf{G}$ singlet scalar coincides with the bound on the $O(N)$ singlet scalar, while the $O(N)$ vector bootstrap bound on the lowest $T$  scalar can be interpreted as the bound on the lowest $\mathbf{G}$ non-singlet scalar appearing in the bootstrap equations of $\mathbf{R}$.

\subsection{$O(N)$ vector bootstrap bounds in the large $N$ limit}
In the large $N$ limit, the $O(N)$ vector crossing equations (\ref{fdcr0},\ref{fdcr}) become
\begin{align}
    \sum_{\cO\in S} \lambda_\cO^2\left(\begin{array}{c}
         0  \\
         E_{\Delta}(z)
    \end{array} \right)+
    \sum_{\cO\in T} \lambda_\cO^2\left(\begin{array}{c}
         F_{\Delta}(z)  \\
         -E_{\Delta}(1-z) 
    \end{array} \right)+\sum_{\cO\in A} \lambda_\cO^2\left(\begin{array}{c}
         -F_{\Delta}(z)  \\
         -E_{\Delta}(1-z)
    \end{array} \right)=0,  \label{fdcr2}
\end{align}
where 
\begin{align}
    E_\Delta(z) &=z^{-2\Delta_\phi}G_\Delta(z),\\ F_\Delta(z)&=E_\Delta(z)-E_\Delta(1-z).
\end{align}
Their bootstrap bound on the lowest non-unit $O(N)$ singlet operator goes to infinity. A solution to such bound is given by the correlation function (\ref{GFF1}-\ref{GFF3}) with $N=\infty$, in which the only $O(N)$ singlet operator is the unit operator, while all the double-trace singlets have vanishing OPE coefficients and are decoupled in the crossing equation.  
Moreover, without extra assumptions on the spectrum, there is no upper bound on the lowest operator in the $T$ or $A$ representation. To show this, let us consider the bootstrap bound  on the scaling dimension of the lowest operator in the $T$ sector, denoted $\Delta_T^*$. A solution to the crossing equation (\ref{fdcr2}) can be constructed as follows.
Given a four point correlator $\langle \phi(x_1)\phi(x_2)\phi(x_3)\phi(x_4)\rangle\sim \cG^*$ which satisfies:
\beq
z^{-2\Delta_\phi}\cG^*(z)-(1-z)^{-2\Delta_\phi}\cG^*(1-z)=0,
\eeq 
the $O(N)$ vector correlation functions
\beq 
\cG^A=\cG^S=\cG^*, 
 ~~\cG^T=0 \label{Fstbd}
\eeq
satisfy the crossing equation (\ref{fdcr2}). 
In this solution the $T$ sector is empty therefore corresponding to an infinity high upper bound $\Delta_T^*=\infty$.
Due to the same logic there is no upper bound on $\Delta_A^*$ either.
Note the unit operator is an indispensable ingredient in the OPEs of the correlation functions $\cG^*$ and $\cG^{S/A}$.
It seems the 1D large 
$N$ bootstrap is too simplified to capture  nontrivial dynamics and is not insightful for higher dimensional bootstrap. However, this is not the case. 

As discussed before, different $O(N)$ representations carry different $\cS$-parity charges, similar to the spin selection rules in higher dimensional bootstrap. With different $\cS$-parity charges it is expected that the spectra in different sectors are notably different. Specifically, in the $O(N)$ vector bootstrap, to bound $\Delta_T^*$, we may expect a non-trivial gap for the lowest operator in the $A$ sector which has opposite $\cS$-parity. In the bootstrap studies of defect CFTs, such gaps can be  justified by the physical spectrum \cite{Gaiotto:2013nva,Gimenez-Grau:2022czc}.

\begin{figure}
\begin{center}
  \includegraphics[width=0.6\linewidth]{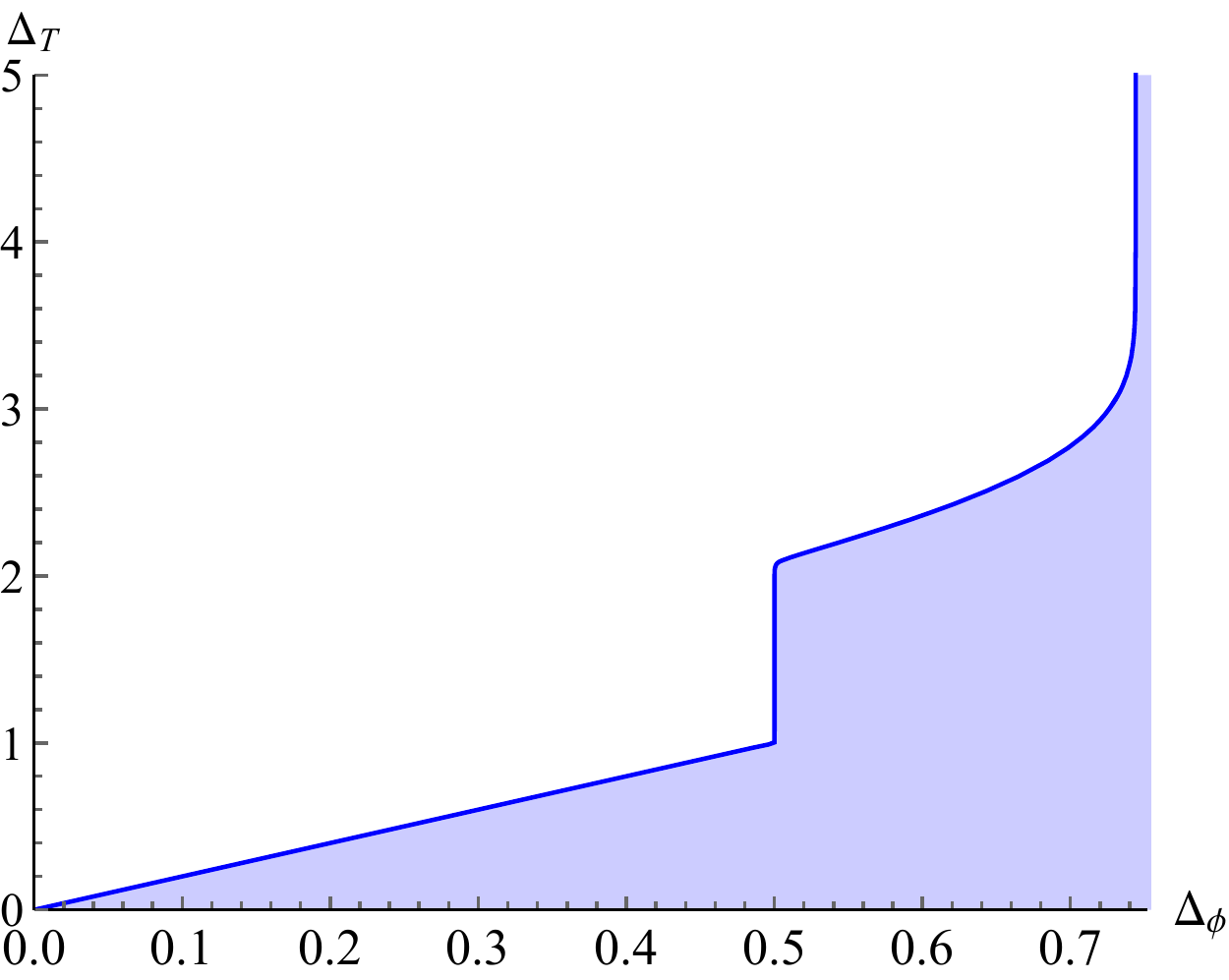}   
\end{center}
\caption{ 1D $O(\infty)$  vector bootstrap bound on the scaling dimension of the lowest operator in the $T$ sector. Gap assumption $\Delta_A>1.0$ in the $A$ sector. $\Lambda=20$.} \label{BD1D}
\end{figure}
\begin{figure}
\includegraphics[width=0.49\linewidth]{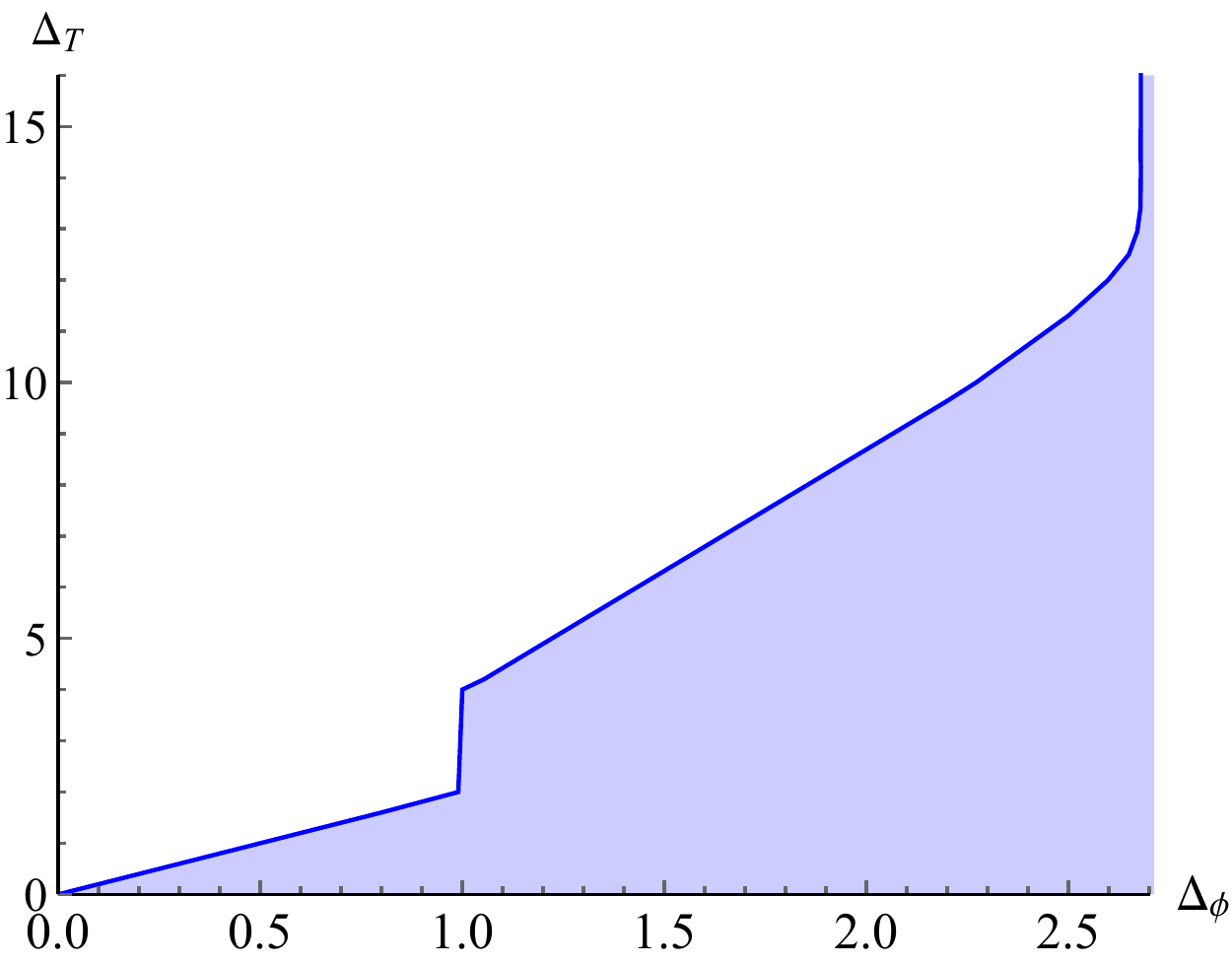}~
\includegraphics[width=0.49\linewidth]{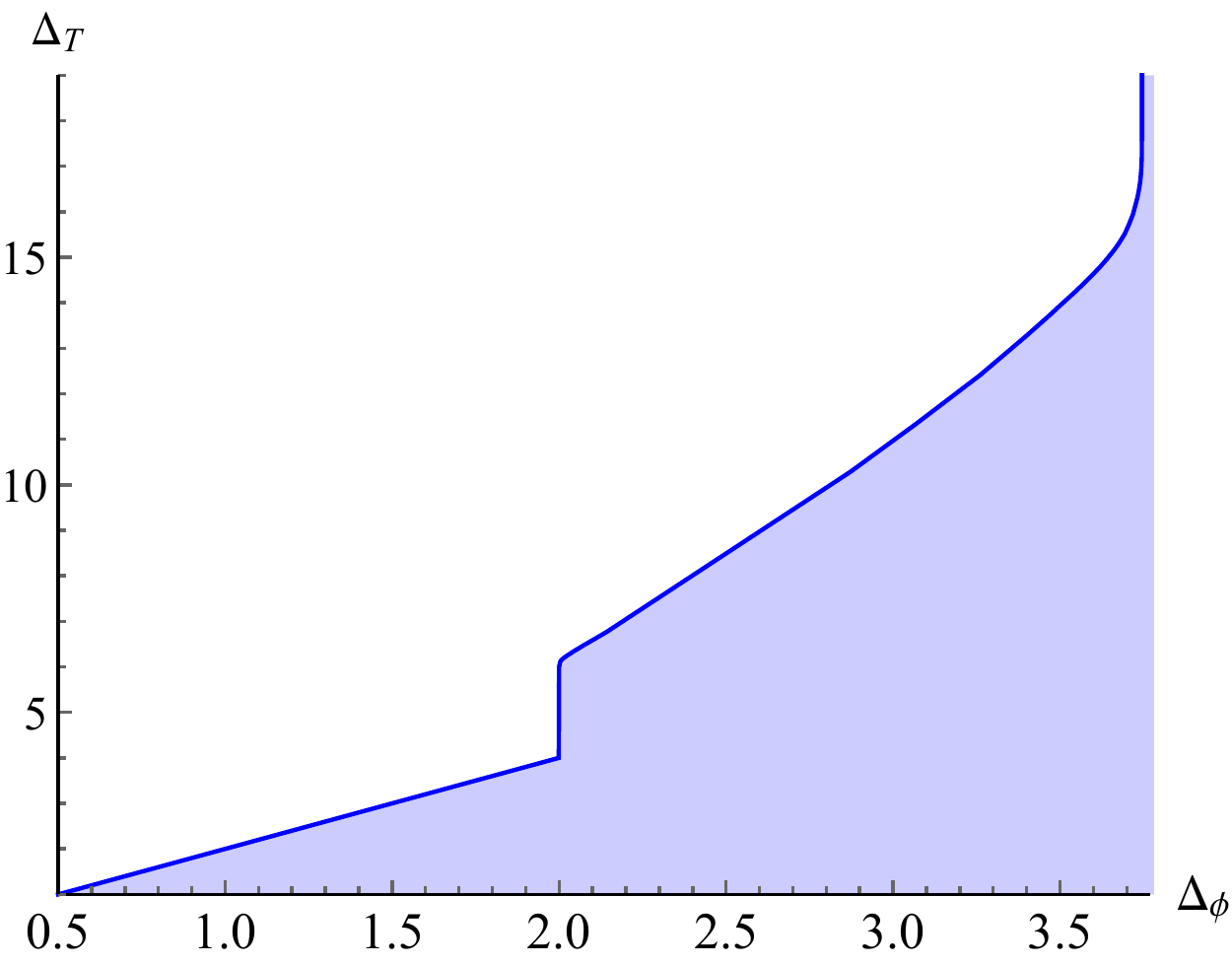}
\caption{2D (left) and 3D (right) $O(\infty)$  vector bootstrap bounds on the scaling dimensions of the lowest scalars in the $T$ sector. No gaps. $\Lambda=31$.} \label{BD23D}
\end{figure}

With a gap assumption on the $A$ sector spectrum, the bootstrap bound on $\Delta_T^*$ can be modified drastically. The result is shown in Fig. \ref{BD1D}, in which we have introduced an assumption that the lowest operator in the $A$ sector satisfies $\Delta_A\geqslant \Delta_c=1$.\footnote{Bootstrap bounds with different gaps $\Delta_c$ are qualitatively similar to Fig. \ref{BD1D}.}  
In the range $\Delta_\phi \in (0,\Delta_c/2)$ the bootstrap bound on $\Delta_T^*$ is given by $\Delta_T^*=2\Delta_\phi$. It is followed by a sharp kink at $\Delta_\phi=\Delta_c/2$, where the bound on $\Delta_T^*$ jumps to $\Delta_T^*=2\Delta_\phi+1$. 
The solution (\ref{Fstbd}) requires a unit operator in the $A$ sector, therefore is excluded by the gap assumption.
The bootstrap bound on $\Delta_T^*$ disappears near $\Delta_\phi=0.744$, which suggest an {\it end of the scalar bootstrap constraints}.\footnote{Note the bound on $\Delta_T^*$ provides the strongest constraint among the scalar bootstrap with global symmetries.}

The 1D $O(\infty)$ vector bootstrap bound is remarkably similar to the higher dimensional $O(\infty)$ vector bootstrap bounds shown in Fig. \ref{BD23D}. It has been known since \cite{Kos:2013tga} that in 3D, the $O(N)$ vector bootstrap bounds show sharp kinks (type I) which are saturated by the 3D critical $O(N)$ vector models.
Moreover, in \cite{Li:2018lyb} the author observed that besides the type I kinks, the 3D $O(N)$ vector bootstrap bounds also show another family of kinks (type II) which approach the free fermion bilinear in the large $N$ limit. The type II kinks appear in general dimensions \cite{Li:2020bnb}, and the kink in Fig. \ref{BD1D} at $\Delta_\phi=\Delta_c/2$ could be considered as their dimensional continuation in 1D. In higher dimensions the type II kinks at finite $N$ are conjectured to be related to the conformal gauge theories, while mixed with the bootstrap bound coincidences due to a positive algebraic structure in the four point crossing equations \cite{Li:2020tsm}. The numerical bootstrap results of the type II kinks are affected by the numerical convergence issue and it is hard to evaluate the CFT data numerically. 

One of the motivations of this work is to develop an analytical functional bootstrap method to study the kinks in the $O(N)$ vector bootstrap bounds and clarify their putative connections to the conformal gauge theories. The higher dimensional bootstrap equations relate to conformal blocks with two cross ratios $z, \bar{z}$ and spins, which make the analytical functional bootstrap more intricate. Here our results suggest that similar bootstrap bounds can also be realized in 1D conformal bootstrap, with a drastically simplified bootstrap setup. Therefore the 1D large $N$ bootstrap can provide a key to unlock the large $N$ analytical functional bootstrap in higher dimensions.

\subsection{Extremal solutions and the simplified bootstrap equation}
\begin{figure}
\includegraphics[width=0.51\linewidth]{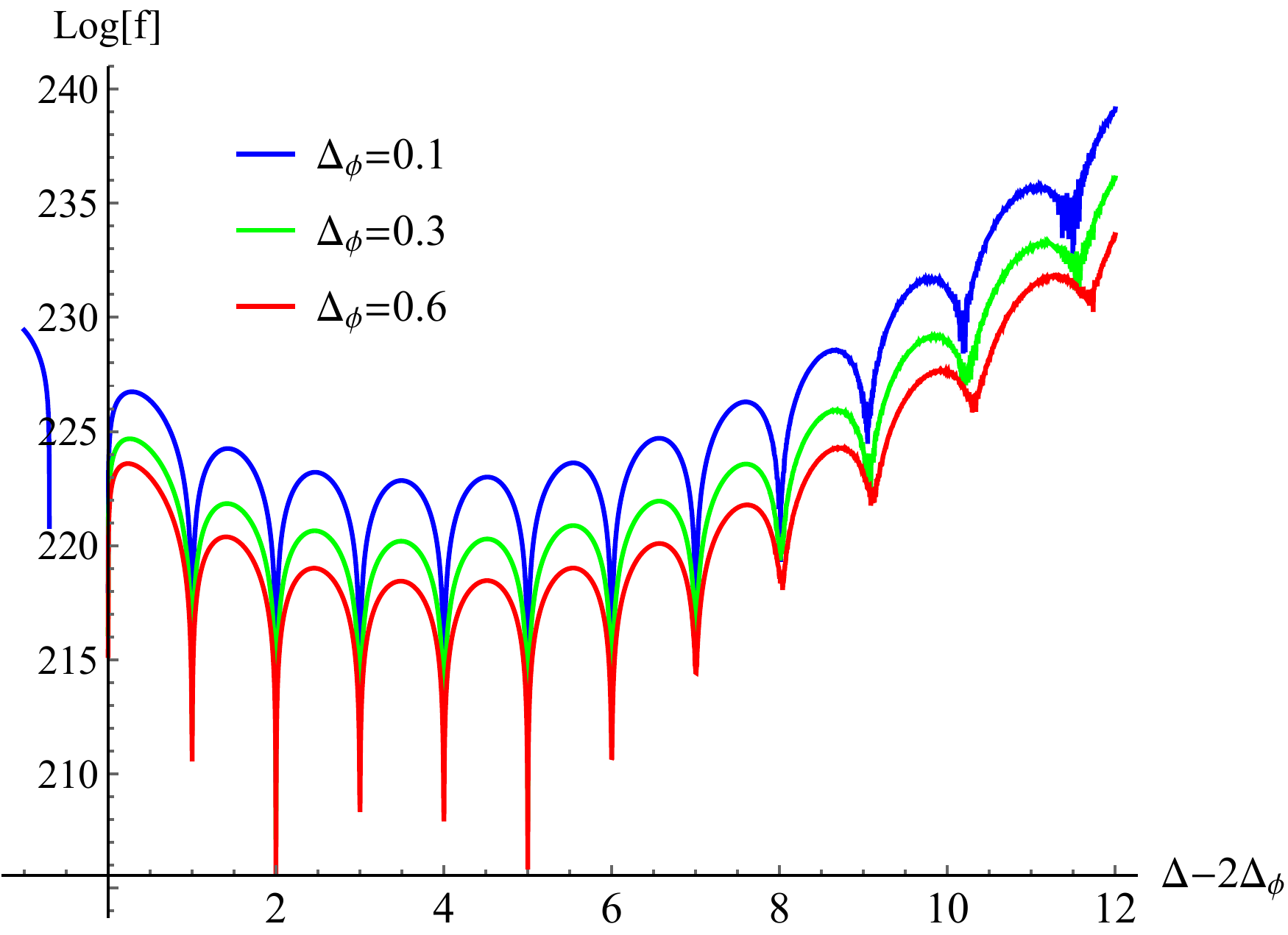}~
\includegraphics[width=0.47\linewidth]{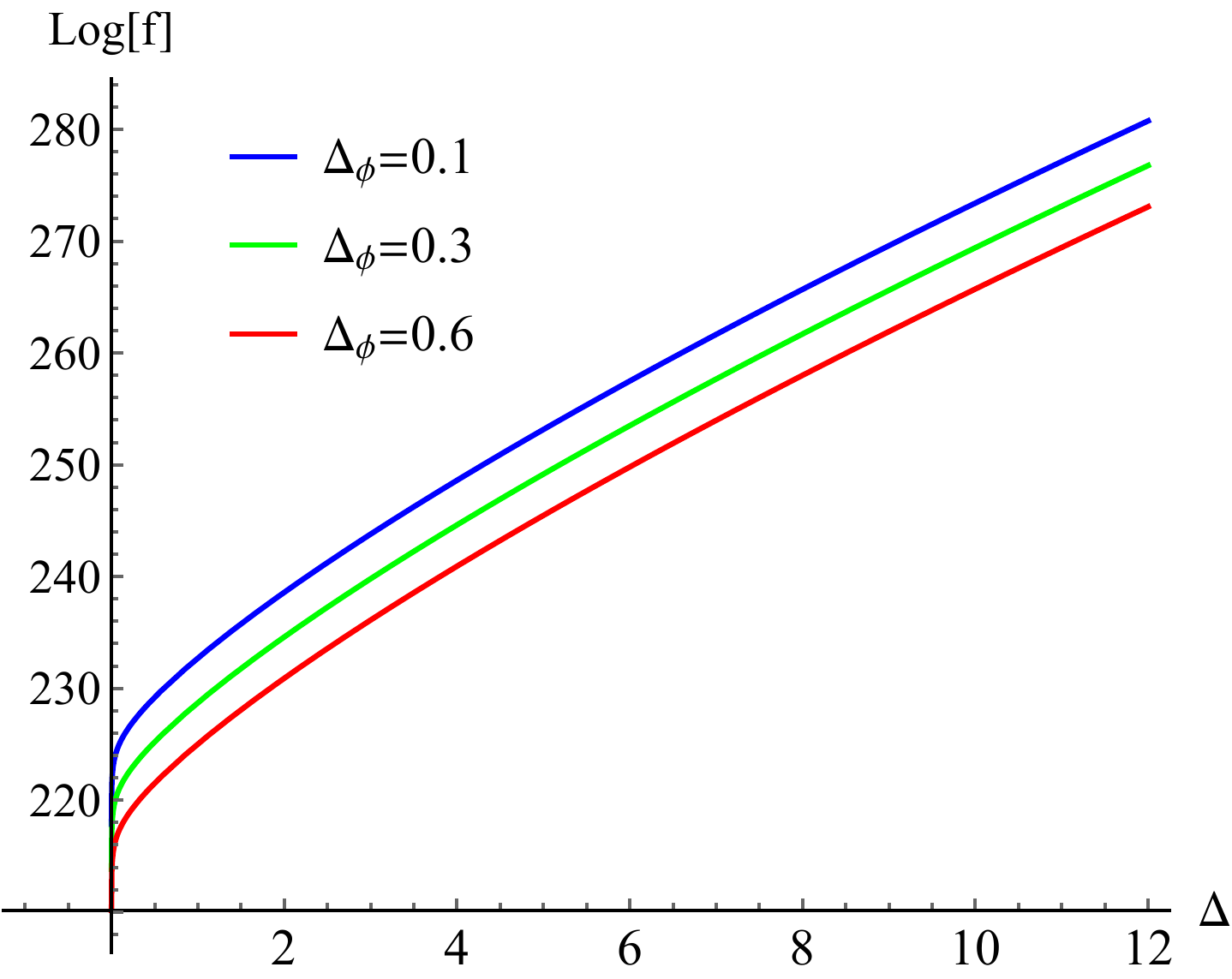}
\caption{Extremal functional spectra in the $O(N)$ $T$ sector (left) and $S$ sector (right). } \label{EFM}
\end{figure}

We focus on the 1D large $N$ bootstrap bound in the range $ \Delta_\phi\in(0,~\Delta_c/2)$. 
Spectrum of the theory saturating the bootstrap bound can be obtained from the extremal functionals \cite{El-Showk:2012vjm}, which are shown in Fig. \ref{EFM} for $\Delta_\phi=0.1,0.3$.
In the $S$ sector the spectrum is trivial with only one first order zero at $\Delta=0$, corresponding to the unit operator. Surprisingly, the extremal functional in the $T$ sector shows a first order zero at $\Delta=2\Delta_\phi$, and double zeros at $\Delta=2\Delta_\phi+n, ~ n\in \mathbb{N}^+$. Therefore the extremal spectrum is not from generalized free boson or fermion alone, but is given by the correlation functions (\ref{GFF1}-\ref{GFF3}) with $|\lambda|<1$. Furthermore, action of the extremal functional in the $A$ sector is the same as that of $T$ sector up to numerical errors! In the $A$ sector, we only introduced the positivity constraint above the gap $\Delta_A\geqslant \Delta_c$, while the extremal solution automatically satisfies the positivity condition down to $\Delta>2\Delta_\phi$!

Let us go back to the $O(\infty)$ vector crossing equation (\ref{fdcr2}) and check what does it mean by two ``almost" identical actions in $T$ and $A$ sectors.
Consider a linear functional $\vec{\alpha}$ for the $O(\infty)$ vector crossing equation (\ref{fdcr2})
\begin{align}
    \vec{\alpha}\cdot \vec{V}_T &\equiv\alpha_1\cdot F_\Delta(z)- \alpha_2\cdot E_\Delta(1-z),  \\
\vec{\alpha}\cdot \vec{V}_A &\equiv -\alpha_1\cdot F_\Delta(z)-  \alpha_2\cdot E_\Delta(1-z).
\end{align}
The observation $\vec{\alpha}\cdot \vec{V}_T=\vec{\alpha}\cdot \vec{V}_A$ suggests $\alpha_1\rightarrow 0\,$! That is to say, to get the upper bound in Fig. \ref{BD1D} for $\Delta_\phi<\Delta_c/2$, the first row in the crossing equation (\ref{fdcr2}) is not necessary! 
The extremely small $\alpha_1$ has been verified in our numerical bootstrap results.

Without the first row of (\ref{fdcr2}), the conformal blocks in the $T$ and $A$ sectors are the same and the positivity constraint in the $A$ sector
\beq 
\vec{\alpha}\cdot \vec{V}_A\geqslant0, ~~~\forall \Delta>\Delta_c,
\eeq
is substituted by the positivity constraint in the $T$ sector
\beq 
\vec{\alpha}\cdot \vec{V}_T\geqslant0, ~~~\forall \Delta>\Delta_T^*,
\eeq
given $\Delta_T^*<\Delta_c$.
While for $\Delta_T^*\geqslant \Delta_c$, the positivity constraints between the two sectors are switched and the bootstrap bound in Fig. \ref{BD1D} suggests the first row of (\ref{fdcr2}) becomes important. The bootstrap constraints have a transition at $\Delta_T^*=\Delta_c$, corresponding to  the jump of the bootstrap bound  at $\Delta_\phi=\Delta_c/2$ in Fig. \ref{BD1D}. We leave a detailed study of the bootstrap bound with $\Delta_\phi\geqslant\Delta_c/2$ for future work.

The correlation functions (\ref{GFF1}-\ref{GFF3}) with different $|\lambda|<1$ have the same spectrum while different OPE coefficients (\ref{OPE1},\ref{OPE2}). The extremal OPE coefficients $c_n^*$ are given by the generalized free boson or fermion theories when $|\lambda|\rightarrow 1$
\beq 
c_n^*=\frac{(2\Delta_\phi)_n^2}{n!(4\Delta_\phi+n-1)_n}. \label{OPE3}
\eeq
We have checked that our bootstrap bounds on the OPE coefficients of low-lying spectrum $\Delta=2\Delta_\phi+n$ are well consistent with (\ref{OPE3}) up to $n=9$. 

To summarize, the $O(\infty)$ vector bootstrap leads to a rather simple crossing equation   
 \beq 
\sum_{\cO\in S}\lambda_{\cO}^2 \;z^{-2\Delta_\phi}\,G_{\Delta}(z) - \sum_{\cO \in T}\lambda_{\cO}^2\, (1-z)^{-2\Delta_\phi} G_{\Delta}(1-z)=0.
\label{simpceq}
\eeq  
Bootstrap bound on $\Delta_T^*$ from above crossing equation is given by $\Delta_T^*=2\Delta_\phi$ for general $\Delta_\phi$, and its extremal spectrum is the same as those in Fig. \ref{EFM}. 
The $O(\infty)$ vector bootstrap equation (\ref{fdcr2}) is reduced to (\ref{simpceq}) in the range $\Delta_\phi<\Delta_c/2$. For $\Delta_\phi\geqslant\Delta_c/2$, the bootstrap bound from (\ref{simpceq}) stays in the line $\Delta_T^*=2\Delta_\phi$, see e.g. the extremal spectrum at $\Delta_\phi=0.6$ in Fig. \ref{EFM}, while the bound from (\ref{fdcr2}) goes differently as the first row in (\ref{fdcr2}) starts to play a role. 
The rest part of this work aims to construct  analytical functionals for the crossing equation (\ref{simpceq}). 

We would like to add comments on  $O(\infty)$ vector bootstrap in higher dimensions \cite{ZL23}. In the range between free boson and free fermion bilinear: $\Delta_\phi\in \left(\frac{D-2}{2}, \, D-1\right)$, the bootstrap bound on $\Delta_T^*$ is also saturated by the generalized free theory and the $O(\infty)$ vector bootstrap equations  are reduced to the higher dimensional form of (\ref{simpceq})
 \begin{align}
   \sum_{\cO\in S}\lambda_{\cO}^2 (z\Bar{z})^{-\Delta_\phi}\,G_{\Delta,\ell}(z,\bar{z}) &- \sum_{\cO \in T}\lambda_{\cO}^2 ((1-z)(1-\Bar{z}))^{-\Delta_\phi}\, G_{\Delta,\ell}(1-z,1-\bar{z}) \nn\\ &- \sum_{\cO \in A}\lambda_{\cO}^2 ((1-z)(1-\Bar{z}))^{-\Delta_\phi}\, G_{\Delta,\ell}(1-z,1-\bar{z})=0,  
 \end{align} 
where $G_{\Delta,\ell}(z,\bar{z})$ are the $SO(D+1,1)$ conformal blocks  \cite{Dolan:2003hv,Dolan:2011dv}.
Considering the close relation between the $O(\infty)$ vector bootstrap in 1D and higher dimensions, the analytical functional for 1D $O(\infty)$ vector bootstrap constructed in this work will be instructive to construct analytical functionals in higher dimensions \cite{ZL23}.

\section{$\SL$ conformal block and total positivity}\label{sec3}
The  1D large $N$ bootstrap  provides an ideal example to decode the underlying mathematical structures of conformal bootstrap. 
Considering there are only few ingredients in the bootstrap crossing equation (\ref{simpceq}), it is expected that the presumed mathematical structures  should be certain properties of the $\SL$ 
conformal block.  In section \ref{sec4} we will construct the analytical functionals for the crossing equation (\ref{simpceq}) and show that the answer to this riddle is {\it total positivity}. In this section we provide a brief explanation of {\it total positivity} and study its relation to the conformal block  $G_\Delta(z)$.

\subsection{Total positivity: definition and theorems} \label{TPdef}

\noindent{\bf Definition.}
A two-variable function $K(x,y)$ defined on $I\times J$ with $I,J\subset \mathbb{R}$ is  totally positive of the order $k$, if for all $1\leqslant m\leqslant k$, and arbitrary ordered variables $x_1<...<x_m, ~y_1<...<y_m,$ $x_i\in I, ~y_j\in J$, the following determinants  are  positive
\bea 
||K(x,y)||_m\equiv K\left(
\begin{array}{ccc}
    x_1, & ... & x_m \\
    y_1, & ... & y_m
\end{array}
\right)=\det \left[\begin{array}{ccc}
   K(x_1,y_1)  & ... & K(x_1,y_m)  \\
   \vdots & & \vdots \\
   K(x_m,y_1)  &...& K(x_m,y_m) 
\end{array}
\right]> 0. \label{TPdf}
\eea 
We are interested in the totally positive functions of the order infinity, which will be assumed implicitly in the following part.
For the finite sets $I, J$, the two-variable functions $K(x,y)$ are reduced to the matrices $K(x,y)\rightarrow K_{i,j}$. In this case, the definition (\ref{TPdf}) for totally positive  matrices becomes that all the minors of the matrix $K$ are positive.

From the definition (\ref{TPdf}), it is straightforward to show following rules for totally positive functions: 
\begin{itemize}
    \item If $g(x)$ and $h(y)$ are positive functions defined on $I$ and $J$, respectively, and $K(x,y)$ is totally positive, then so is the function $g(x)K(x,y)h(y)$.
    \item If $g(x)\in I$ and $h(y)\in J$ are defined on $x\in U$ and $y\in V$, and monotone in the same direction, and if $K(x,y)$ is totally positive on $I\times J$, then the function $K(g(x),h(y))$ is  totally positive on $U\times V$.
\end{itemize}
 
An important tool to study total positivity is the so-called ``basic composition formula". It shows how to construct a new totally positive function from two such functions and provides a powerful method to prove total positivity of certain functions.
\vspace{3mm}

\noindent{\bf Basic composition formula.} Let  $K,L,M$ be two-variable functions which satisfy
\begin{equation}
  M(x,y)=\int K(x,z) L(z,y)d \sigma(z),  \label{CBf}
\end{equation} 
where $\sigma(z)$ is a $\sigma$-finite measure and the integral converges absolutely, then the basic composition formula suggests
\bea
M\left(
\begin{array}{ccc}
    x_1, & ... & x_m \\
    y_1, & ... & y_m
\end{array}
\right)=&
\nn\\
\idotsint\limits_{z_1<\dots<z_m} &
K\left(
\begin{array}{ccc}
    x_1, & ... & x_m \\
    z_1, & ... & z_m
\end{array}
\right)
L\left(
\begin{array}{ccc}
    z_1, & ... & z_m \\
    y_1, & ... & y_m
\end{array}
\right)d\sigma(z_1)\dots d\sigma(z_m). \label{BCf}
\eea
A proof of this formula is sketched in \cite{erdelyi_1970}.

The convolution (\ref{CBf}) of two kernels $K(x,z)$ and $L(z,y)$  can be considered as a continuous version of the standard matrix product, then above basic composition formula (\ref{BCf}) is an extension of the Cauchy-Binet formula in matrix multiplication which expands subdeterminants of $M_{ij}$ in terms of those of $K_{im}$ and $L_{mj}$. The basic composition formula  (\ref{BCf}) directly leads to following theorem.
\vspace{3mm}
 
\noindent{\bf Theorem.} 
The convolution (\ref{CBf}) of two totally positive kernels is also totally positive.

\vspace{3mm}
\noindent{\bf Variation Diminishing Property.} 
Consider a function $f: I\rightarrow\mathbb{R}$, where $I\subset \mathbb{R}$. The {\it number of sign changes} of $f$ on $I$, denoted $S_I^+(f)$, is defined as the maximum number of sign changes in a finite sequence $\{f(x_1), f(x_2),\dots, f(x_m)\}$, $x_i\in I$, $x_1<\dots<x_m$.\footnote{The zeros in the sequence  are  discarded when counting the number of sign changes.}
An important property of the totally positive functions is given by \cite{erdelyi_1970}:  
\vspace{3mm}

\noindent{\bf Theorem.} For $I, J\subset \mathbb{R}$, consider a totally positive kernel $K: I\times J\rightarrow \mathbb{R}$ which is Borel-measurable. Let $\sigma(y)$ be a regular $\sigma$-finite measure on $J$ and $f: J\rightarrow \mathbb{R}$ be a bounded and Borel-measurable function on $J$, so that the convolution of $f(y)$ converges absolutely
\begin{equation}
    g(x)=\int_J K(x,y)f(y)d\sigma(y). \label{VDP}
\end{equation}
Then the number of sign changes of $g(x)$ on $I$ is not larger than that of $f(y)$ on $J$:
\begin{equation}
    S_I^+(g)\leqslant S_J^+(f). \label{VDP1}
\end{equation}
Moreover, if $S_I^+(g)= S_J^+(f)$, then the two functions $f(y)$ and $g(x)$ should have the same arrangement of signs.
\vspace{3mm}

The {\it variation diminishing property} of totally positive functions will play a critical role to construct analytical functionals of $1D$ large $N$ bootstrap.

\subsection{Total positivity of the Gauss hypergeometric function} \label{TP2F1}
The Gauss hypergeometric function $_2F_1(\Delta,\Delta,2\Delta,z)$ appears in the $\SL$ conformal block $G_\Delta(z)$.  There is numerical evidence indicating that the function $_2F_1(\Delta,\Delta,2\Delta,z)$ is indeed totally positive in the region $z\in (0,1), ~\Delta>0$ \cite{Arkani-Hamed:2018ign}. We have also numerically verified the total positivity of this function using a large set of data.
While it is hard to obtain a complete proof for the total positivity of $_2F_1(\Delta,\Delta,2\Delta,z)$, we can get promising evidence for this observation beyond the numerical checks.

\subsubsection*{Total positivity of $_2F_1(\Delta,\Delta,2\Delta,z)$ in the large $\Delta$ limit}
In the large $\Delta$
limit, the hypergeometric function $_2F_1(\Delta,\Delta,2\Delta,z)$ has a much simpler asymptotic form, 
for which the total positivity can be proved easily.
Let us consider the integral formula of the hypergeometric function 
\begin{equation}
_2F_1(\Delta,\Delta,2\Delta,z)=\frac{1}{B(\Delta,\Delta)}\int_0^1 x^{\Delta-1}(1-x)^{\Delta-1}(1-zx)^{-\Delta}dx, \label{Euler2F1}
\end{equation}
where $B(\Delta,\Delta)=\frac{\Gamma(\Delta)^2}{\Gamma(2\Delta)}$ is the Euler Beta function.
In the large $\Delta$ limit above integration can be solved using the method of steepest descent:
\begin{equation}
\int_0^1 x^{\Delta-1}(1-x)^{\Delta-1}(1-zx)^{-\Delta}dx=\int_0^1 \frac{1}{x (1-x)}e^{-\Delta  \log \left[\frac{1-x z}{x (1-x)}\right]}dx,
\end{equation}
which has a single stationary point $x=\frac{1-\sqrt{1-z}}{z}$ in the region $x\in(0,1)$. Then the integration (\ref{Euler2F1}) is approximately given by
\begin{equation}
_2F_1(\Delta,\Delta,2\Delta,z)|_{\Delta\rightarrow\infty}\approx  \frac{1}{B(\Delta,\Delta)} \sqrt{\frac{\pi }{\Delta}} (1-z)^{-\frac{1}{4}} \left(1+\sqrt{1-z}\right)^{1-2 \Delta}. \label{App2F1}
\end{equation}
We find above approximation is reasonably good even for $\Delta=5$.

It is straightforward to prove the total positivity of the right hand side of (\ref{App2F1}). Since the positive factors  depending solely on $z$ or $\Delta$ have no effect on the total positivity, the only relevant factor in the approximated formula is
\begin{equation}
    \left(1+\sqrt{1-z}\right)^{-2 \Delta}=\rho(z)^{2\Delta}, \label{rho1}
\end{equation}
where $\rho(z)=\left(1+\sqrt{1-z}\right)^{-1}$ is a monotone increasing function in $z\in(0,1)$. Therefore the asymptotic formula (\ref{App2F1}) has the same total positivity as the function $z^\Delta$, which has been proved in Appendix \ref{TPEx1}.

\subsubsection*{A sufficient condition for the total positivity of $_2F_1(\Delta,\Delta,2\Delta,z)$}
Both the large $\Delta$ approximation and numerical tests with small $\Delta$'s suggest the hypergeometric function $_2F_1(\Delta,\Delta,2\Delta,z)$ is totally positive. Here we discuss a sufficient condition which, if true, can prove the totally positivity of $_2F_1(\Delta,\Delta,2\Delta,z)$ for general $\Delta>0$.

The hypergeometric function has a series expansion
\begin{equation}
_2F_1(\Delta,\Delta,2\Delta,z)=\sum_{i=0}^\infty \frac{(\Delta)_i^2}{(2\Delta)_i} \frac{z^i}{i!},
\end{equation}
where $(a)_i$ is the Pochhammer symbol. Above expansion can be considered as a convolution of $K(\Delta,i)\equiv (\Delta)_i^2/(2\Delta)_i$ and $f(i,z)\equiv z^i/i!$  with a discrete $\sigma$-measure in (\ref{CBf}). Therefore according to the basic composition formula (\ref{BCf}), the hypergeometric function
$_2F_1(\Delta,\Delta,2\Delta,z)$ is totally positive if both of the two functions $K(\Delta,i)$ and $f(i,z)$ are totally positive.  The function $z^i$ has been shown to be totally positive. 

The total positivity of the function $K(\Delta,i)$ requires 
\bea 
||K(\Delta,i)||_m=K\left(
\begin{array}{ccc}
    \Delta_1, & ... & \Delta_m \\
    i_1, & ... & i_m
\end{array}
\right)=\det \left[\begin{array}{ccc}
   \frac{(\Delta_1)_{i_1}^2}{(2\Delta_1)_{i_1}}  & ... & \frac{(\Delta_m)_{i_1}^2}{(2\Delta_m)_{i_1}}  \\
   \vdots & & \vdots \\
   \frac{(\Delta_1)_{i_m}^2}{(2\Delta_1)_{i_m}}  &...& \frac{(\Delta_m)_{i_m}^2}{(2\Delta_m)_{i_m}} 
\end{array}
\right]> 0, 
\eea 
with $0\leqslant \Delta_1<\dots<\Delta_m, ~ 0\leqslant i_1<\dots<i_m$, $\Delta_k\in\mathbb{R},~ i_k\in \mathbb{N}$ for any integer $m$. A compact formula for above determinants with general $m$ is not known. Here we show for small $m$, above determinants are indeed positive.

Consider the determinant $||K(\Delta,i)||_{m=2}$ for  general $\Delta_k$ and $i_k$ in the domain of definition
\begin{align}
    ||K(\Delta,i)||_{2} &= \det \left[\begin{array}{cc}
   \frac{(\Delta_1)_{i_1}^2}{(2\Delta_1)_{i_1}}  &   \frac{(\Delta_2)_{i_1}^2}{(2\Delta_2)_{i_1}}  \\
   \frac{(\Delta_1)_{i_2}^2}{(2\Delta_1)_{i_2}}  &\frac{(\Delta_2)_{i_2}^2}{(2\Delta_2)_{i_2}} 
\end{array}
\right] \nn\\  
&=
\frac{(\Delta_2)_{i_1}^2 (\Delta_1)_{i_2}^2}{(2\Delta_2)_{i_1}(2\Delta_1)_{i_2}} 
\left(\;\prod\limits_{k=0}^{i_2-i_1-1} \frac{(\Delta_2+i_1+k)^2 (2\Delta_1+i_1+k)}{(\Delta_1+i_1+k)^2 (2\Delta_2+i_1+k)}  -1 \right). \label{DET2}
\end{align}
For each term in the product with $k\geqslant0$, we have
\begin{align}
    \left(\Delta _2+i_1+k\right){}^2\left(2 \Delta _1+i_1+k\right)& -\left(\Delta _1+i_1+k\right){}^2\left(2 \Delta _2+i_1+k\right) = \nn\\
    & \hspace{-1cm} \left(\Delta _2-\Delta _1\right) \left(2 \Delta _2 \Delta _1+\Delta _1 i_1+\Delta _2 i_1+\Delta _1 k+\Delta _2 k\right)>0,
\end{align}
and consequently
\begin{equation}
    \frac{(\Delta_2+i_1+k)^2 (2\Delta_1+i_1+k)}{(\Delta_1+i_1+k)^2 (2\Delta_2+i_1+k)}>1.
\end{equation}
Therefore the right hand side of (\ref{DET2}) is positive.  

With higher $m$'s the determinant formula $||K(\Delta,i)||_{m}$ is too complicated for a general study. By choosing a specific set of $i_k$'s one can evaluate the determinants explicitly. 
For instances, taking $i_k=k$, the determinants $||K(\Delta,k)||_m$ are given by 
\begin{align}
||K(\Delta,k)||_{m=3} =& \label{fm3}
 \\
&\hspace{-1.5cm} \left(\Delta _2-\Delta _1\right) \left(\Delta _3-\Delta _1\right) \left(\Delta _3-\Delta _2\right)\Delta _1 \Delta _2 \Delta _3 
   \frac{  \left(\Delta _1 \Delta _2+\Delta _3 \Delta _2+\Delta _1 \Delta _3+2 \Delta _1 \Delta _2 \Delta _3\right) }{16 \left(2 \Delta _1+1\right) \left(2 \Delta _2+1\right) \left(2 \Delta _3+1\right)}  \nn
\end{align}
for $m=3$
and 
\begin{align}
||K(\Delta,k)||_{m=4} = & \label{fm4}\\
 &\hspace{-1cm} \left(\Delta _2-\Delta _1\right) \left(\Delta _3-\Delta _1\right) \left(\Delta _3-\Delta _2\right) \left(\Delta _4-\Delta _1\right) \left(\Delta _4-\Delta _2\right) \left(\Delta _4-\Delta _3\right) \Delta _1 \Delta _2 \Delta _3 \Delta _4  \nn\\
 &\times\left(3 \Delta _1 \Delta _2 \Delta _3 \left(9 \Delta _3+\Delta _1 \left(2 \Delta _2+3\right) \left(2 \Delta _3+3\right)+\Delta _2 \left(6 \Delta _3+9\right)+13\right)+\right.\nn\\
& \hspace{-3cm}\left(9 \Delta _3+\Delta _1 \left(2 \Delta _2+3\right) \left(2 \Delta _3+3\right)+\Delta _2 \left(6 \Delta _3+9\right)+13\right) \left(3 \Delta _2 \Delta _3+\Delta _1 \left(3 \Delta _3+\Delta _2 \left(8 \Delta _3+3\right)\right)\right) \Delta _4 \nn\\
& \left.+\left(2 \Delta _1+3\right) \left(2 \Delta _2+3\right) \left(2 \Delta _3+3\right) \left(\Delta _2 \Delta _3+\Delta _1 \left(\Delta _3+\Delta _2 \left(2 \Delta _3+1\right)\right)\right) \Delta _4^2\right) \nn \\
 &\hspace{-2.3cm} /64 \left(2 \Delta _1+1\right) \left(2 \Delta _1+3\right) \left(2 \Delta _2+1\right) \left(2 \Delta _2+3\right) \left(2 \Delta _3+1\right) \left(2 \Delta _3+3\right) \left(2 \Delta _4+1\right) \left(2 \Delta _4+3\right) \nn
\end{align}
for $m=4$,
both of which are obviously positive for ordered $\Delta_i$'s. 
In all similar checks we find the results are well consistent with the total positivity. 
We conjecture this function is totally positive at infinity order for general $\Delta\geqslant0$.

\subsection{Total positivity of the 1D $\SL$ conformal block}
Now we study the total  positivity of the fundamental ingredient 
in 1D conformal bootstrap, the $\SL$ conformal block $G_\Delta(z) =z^\Delta 
\, _2F_1(\Delta,\Delta,2\Delta,z)$, which is a product (but not convolution) of two totally positive factors. 
However, it is {\it not} guaranteed that the product of two totally positive functions is also totally positive, and indeed, the function $G_\Delta(z)$ loses its total positivity in the region with  small $\Delta_i$. This surprising fact was firstly observed in \cite{Arkani-Hamed:2018ign}.\footnote{The author would like to thank Nima Arkani-Hamed for  the inspiring discussion on this problem.} 
Here we study the total positivity of $G_\Delta(z)$ from different aspects.

\subsubsection*{Total positivity of $G_\Delta(z)$ in the large $\Delta$ limit}
Using the asymptotic formula (\ref{App2F1}) of the Gauss hypergeometric function, the large $\Delta$ limit of the 1D conformal block is given by
\begin{equation}
    G_\Delta(z)|_{\Delta\rightarrow \infty} \approx
    \frac{1}{B(\Delta,\Delta)} \sqrt{\frac{\pi }{\Delta}} (1-z)^{-\frac{1}{4}}\, z^\Delta \left(1+\sqrt{1-z}\right)^{1-2 \Delta}. \label{AppCB}
\end{equation}
The total positivity of above formula is determined by the factors depending on both $z$ and $\Delta$:
 \begin{equation}
    z^\Delta \left(1+\sqrt{1-z}\right)^{-2 \Delta}=\Tilde{\rho}(z)^{\Delta}, \label{Slrho}
\end{equation}
where $\Tilde{\rho}(z)=z\,(1+\sqrt{1-z})^{-2}$,\footnote{Interestingly, the variable $\Tilde{\rho}$ in (\ref{Slrho}) is just the variable $\rho(z)$ in \cite{Hogervorst:2013sma} motivated by different reasons. } like $\rho(z)$ in (\ref{rho1}), is a monotone increasing function in $z\in(0,1)$. Thus the 1D conformal block function is totally positive for sufficiently large $\Delta$. However, for small $\Delta$, the large $\Delta$ approximation (\ref{AppCB}) fails and it cannot say anything about the total positivity of $G_{\Delta}(z)$ with small $\Delta$.

\subsubsection*{A ``Fixed point" of the 1D conformal block $G_{\Delta}(z)$}
We show an interesting property of the conformal block $G_\Delta(z)$, though its physical correspondence is not clear yet.

\begin{figure}
\includegraphics[width=0.48\linewidth]{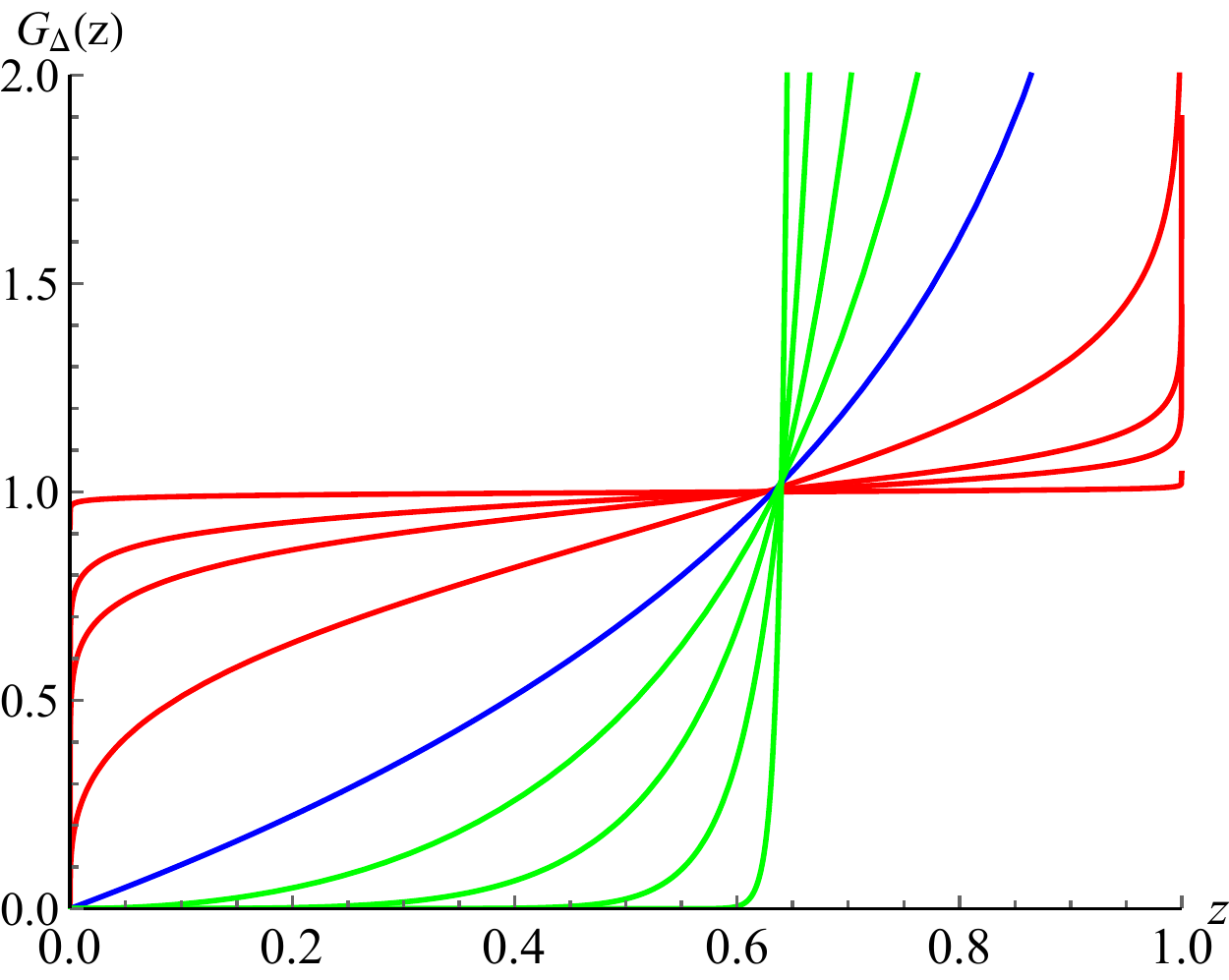}~~~
\includegraphics[width=0.48\linewidth]{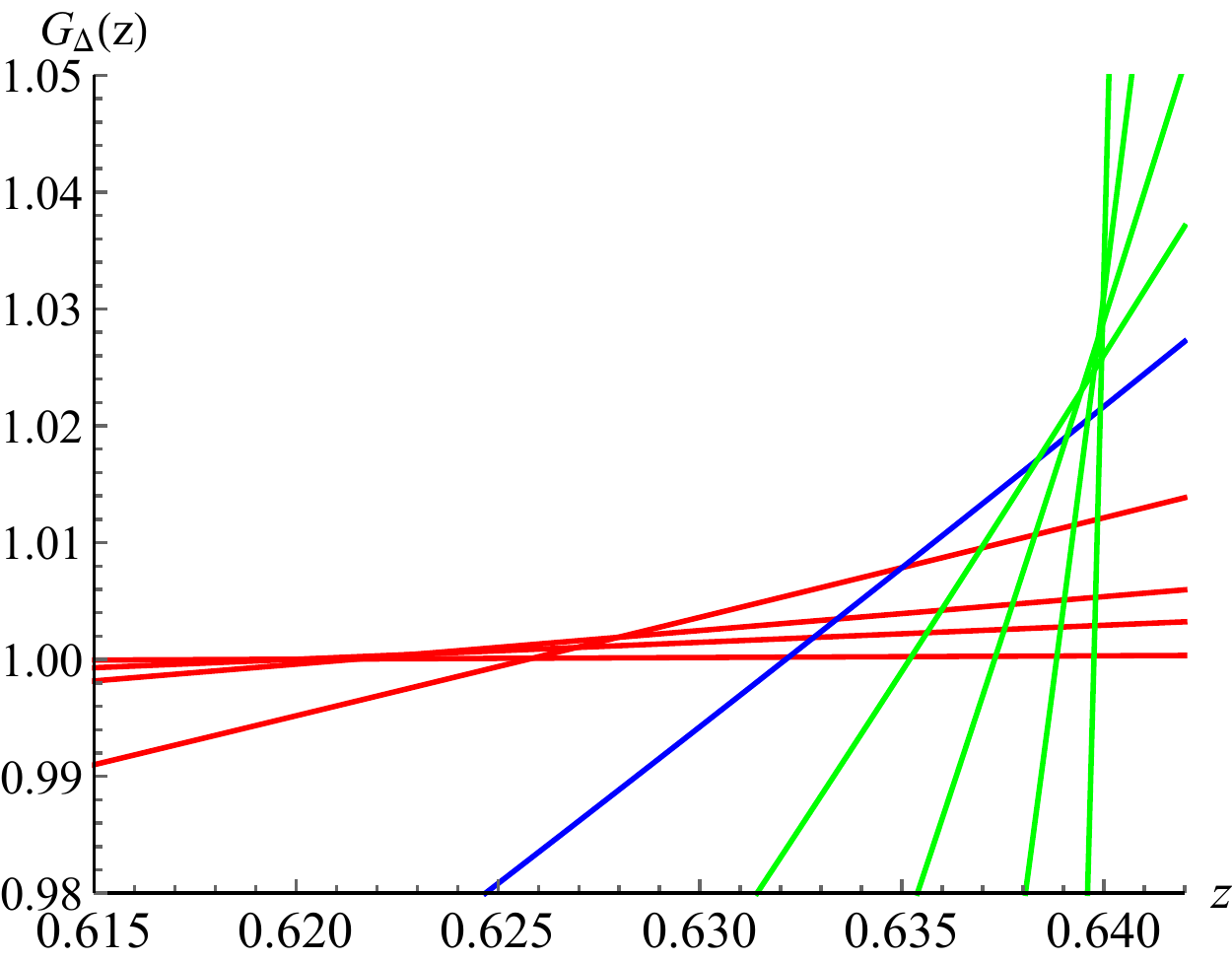}
\caption{Plots for the conformal block functions $G_{\Delta}(z)$ with $\Delta=0.005,0.05,0.1,0.3$ (red curves), $\Delta=1$ (blue curve) and $\Delta=2,4,10,50$ (Green curves). } \label{FPCB}
\end{figure}

The conformal blocks $G_\Delta(z)$ with different $\Delta$'s are plotted in Fig. \ref{FPCB}. A surprising fact is that  all these functions intersect near $\Delta\in( 0.62,0.64)$ with $G_\Delta(z)\simeq 1$. This tiny intersection region looks like a ``fixed point" of the conformal block $G_\Delta(z)$,  besides another trivial ``fixed point" at $z=0$.   Why?

Let us first consider the large $\Delta$ approximation of $G_\Delta(z)$ (\ref{AppCB}).  The dominating part of $G_\Delta(z)$ in this limit is 
\begin{equation}
   G_\Delta(z)|_{\Delta\rightarrow\infty}\sim \frac{1}{B(\Delta,\Delta)}\left(\frac{z}{(1+\sqrt{1-z})^2}\right)^\Delta\approx 4^\Delta\left(\frac{1-\sqrt{1-z}}{1+\sqrt{1-z}}\right)^\Delta. \label{GLDta}
\end{equation}
Here we have used the Stirling's formula for the Gamma function which gives $B(\Delta,\Delta)\sim 4^{-\Delta}$. From (\ref{GLDta}) it is clear that in the large $\Delta$ limit, the equation $G_\Delta(z)=1$, or $\log (G_\Delta(z))=0$ has a $\Delta$-independent solution at
$z=0.64$. Contributions from extra factors are exponentially suppressed.

Then let us go to the small $\Delta$ limit. With a small $\Delta$ the Gauss hypergeometric function is simplified to \begin{equation}
_2F_1(\Delta,\Delta,2\Delta,z)|_{\Delta\ll1}\approx 1-\frac{\Delta}{2} \log(1-z)+O(\Delta^3)    \label{HPGapp}
\end{equation}
and the conformal block function $ G_\Delta(z)$ becomes
\begin{equation}
    G_\Delta(z)|_{\Delta\ll1}  \approx 1+\Delta\,\log\left(\frac{z}{\sqrt{1-z}}\right) +O(\Delta^2), \label{AppGz}
\end{equation}
in which the equation $G_\Delta(z)=1$ is solved by $z=(\sqrt{5}-1)/2\approx 0.618$.

So both in the large and small $\Delta$ limits, the equation $ G_\Delta(z)=1$ has a solution independent of $\Delta$. The solution {\it walks} slowly from $z\approx0.618$ near $\Delta=0$ to $z=0.64$ near $\Delta=\infty$.
Such a ``fixed point" shows an interesting interplay between the factors $z^\Delta$ and the hypergeometric function $_2F_1(\Delta,\Delta,2\Delta,z)$ in $G_\Delta(z)$. As will be shown below, the factor $z^\Delta$ also changes the total positivity of $G_\Delta(z)$ with small $\Delta_i$.

\subsubsection*{Loss of total positivity of $ G_\Delta(z)$ with small $\Delta$s}
\begin{figure}
\begin{center}
\includegraphics[width=0.5\linewidth]{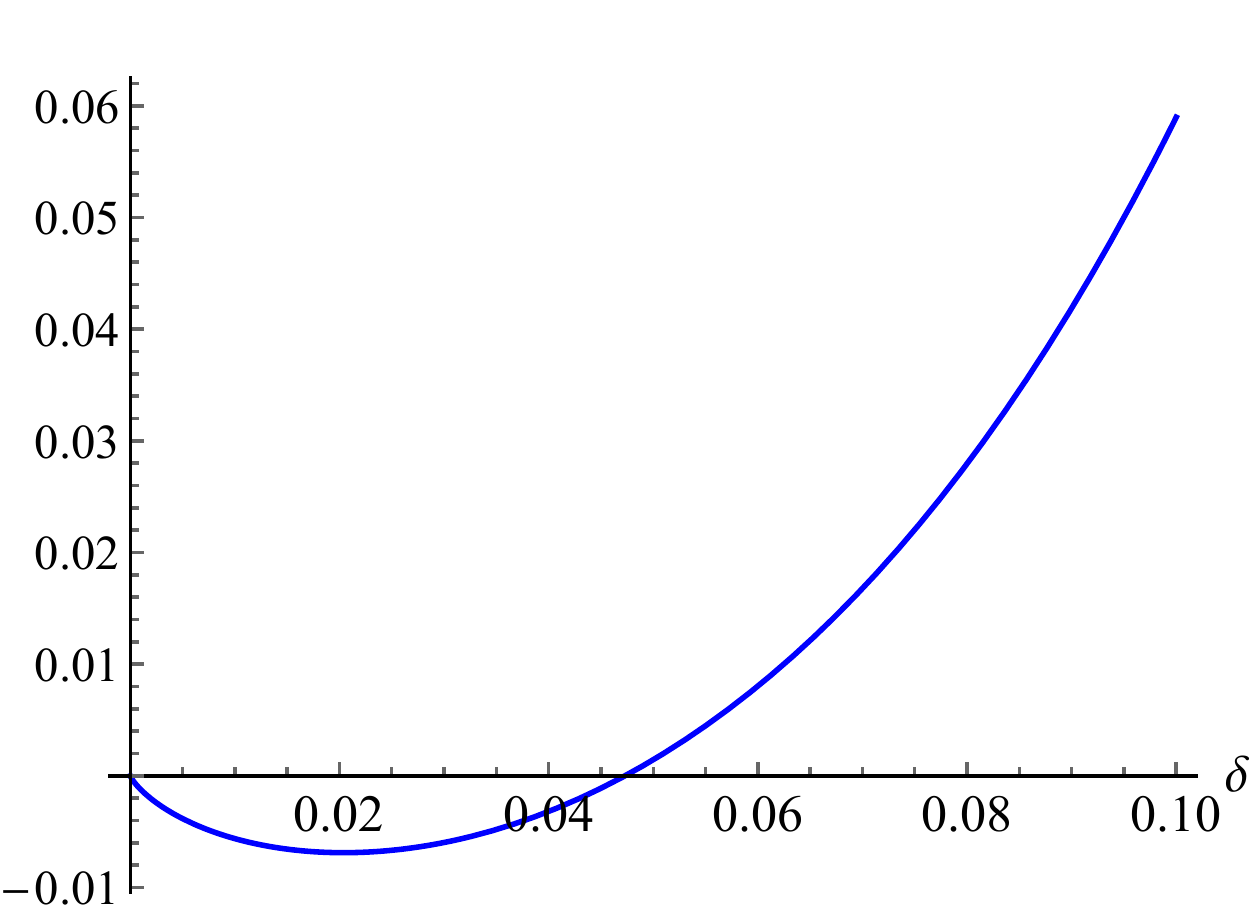} 
\end{center}
\caption{The $z$-dependent factor in (\ref{DtGm3}) is negative in a small range $0<\delta<0.047$. } \label{Ngz}
\end{figure}

We show the total positivity  is violated by the 1D conformal block $G_\Delta(z)$  with  small  $\Delta_i\ll 1$ at the order 3. 
For sufficiently small $\Delta$ it is convenient to take the lower order expansion (\ref{AppGz}) of $G_\Delta(z)$. Up to the order $\Delta^2$, it is given by 
\begin{equation}
    G_\Delta(z)|_{\Delta\ll1}\approx1+\Delta\,\log\left(\frac{z}{\sqrt{1-z}}\right) -\Delta^2\log(z)\tanh^{-1}(1-2z)+O(\Delta^3),
\end{equation}
Let us consider the determinant of 
$G_\Delta(z)$ at the third order
\begin{align}
     ||G_{\Delta}(z)||_3 &=\det\left[\begin{array}{ccc}
   G_{\Delta_1}(z_1)  &   G_{\Delta_1}(z_2) & G_{\Delta_1}(z_3)  \\
  G_{\Delta_2}(z_1)  &   G_{\Delta_2}(z_2) & G_{\Delta_2}(z_3)  \\
  G_{\Delta_3}(z_1)  &   G_{\Delta_3}(z_2) & G_{\Delta_3}(z_3) 
\end{array} 
\right]   \label{DtGm3} \\
&=\frac{\left(\Delta _2-\Delta _1\right) \left(\Delta _3-\Delta _1\right) \left(\Delta _3-\Delta _2\right)}{4}
\left(\log  \left(z_1\right)\log\left(\frac{z_1}{1-z_1}\right)  \log \left(\frac{z_3^2 \left(1-z_2\right)}{z_2^2\left(1-z_3\right) }\right) \right.\nn\\
& \hspace{-1.8cm}\left.+\log \left(z_3\right) \log \left(\frac{z_3}{1-z_3}\right) \log \left(\frac{z_2^2 \left(1-z_1\right)}{ z_1^2\left(1-z_2\right)}\right) 
+\log \left(z_2\right) \log \left(\frac{z_2}{1-z_2}\right)  \log \left(\frac{z_1^2 \left(1-z_3\right)}{z_3^2\left(1-z_1\right) }\right)\right), \nn
\end{align}
in which the $\Delta$ factors are positive for the ordered $\Delta_i$. However, the $z_i$-dependent factor is not definitely positive. Considering $z_i=1+(i-4)\delta$ with a small variable $\delta$, at the leading order the $z$-dependent factor in (\ref{DtGm3}) is
\begin{equation}
    \left( \log \frac{4}{3}\, \log \delta +\log 2 \,\log 3\right)\,\delta +O (\delta^2), \label{negativept}
\end{equation}
which is negative for $0<\delta<2^{-\frac{\log 3}{\log  \frac{4}{3} }}$.  
A nonperturbative plot of the whole $z$-dependent factor in (\ref{DtGm3}) is shown in Fig. \ref{Ngz}. This confirms that the total positivity is violated by the function $G_\Delta(z)$ with small $\Delta_i$ and $1-z$.\footnote{In contrast, the small $\Delta$ expansion of the Gauss hypergeometric function $\,_2F_1(\Delta,\Delta,2\Delta, z)$ does not generate such negative determinants and is always totally positive.}

Using the same approach one can compute the determinant at the forth order $||G_\Delta(z)||_{4}$, and its limit with small $\Delta$ and $1-z$ is actually positive
\begin{equation}
    ||G_\Delta(z)||_{4}\approx0.024\, \delta^2 \,\prod\limits_{i<j}\left(\Delta _j-\Delta _i\right).
\end{equation}
Non-positive determinants appear again at the fifth order. The $\Delta$-expansion of $G_\Delta(z)$ at the order 
$O(\Delta^4)$ is rather complicated and similar analytical results for $||G_\Delta(z)||_{5}$ are not available yet. Numerically one can show that for $\Delta_i=0.1*i, ~z_j=1+0.001(j-6)$, $||G_\Delta(z)||_{5}\approx-8.7\times10^{-26}$. 

\subsubsection*{Critical value $\Delta_{\textrm{TP}}^*$ for the total positivity of $G_\Delta(z)$}
Above computations show the total positivity of $G_\Delta(z)$ is violated with small $\Delta$ and $1-z$. In contrast, in the large $\Delta$ limit   the 1D conformal block is totally positive. The non-positive determinants $||G_\Delta(z)||_m$ should disappear above certain threshold value $\Delta>\Delta_{\textrm{TP}}^*$.
To determine $\Delta_{\textrm{TP}}^*$ is of critical importance for conformal bootstrap study. Moreover, it uncovers a  surprising mathematical structure of the 1D conformal block.

We numerically evaluate the  determinants $||G_\Delta(z)||_m$ using the exact formula  of $G_\Delta(z)$ (\ref{1DCB}). We firstly adopt evenly distributed data to compute $||G_\Delta(z)||_m$ and the determinants with general data will be studied later:
\begin{equation}
    \Delta_i=\Delta_{\textrm{TP}}*i, ~~~ z_j=1+\epsilon\, (j-m-1),   ~~~~1\leqslant i,j\leqslant m. \label{Dataset}
\end{equation}
Numerical results show that the negative determinants appear for odd $m$  with 
 $\Delta_{\textrm{TP}}<\Delta_{\textrm{TP}}^c$ and  $\epsilon<\epsilon^c$, in which the threshold values  depend on $m$.  The threshold value $\Delta_{\textrm{TP}}^c$ for $||G_\Delta(z)||_m>0$ decreases with increasing $m$,  and the maximum estimation $\Delta_{\textrm{TP}}^c\approx 0.1627$ is obtained with $m=3$, see Table \ref{DetGm}.

\begin{table}
\caption{Smallest  $\Delta_{\textrm{TP}}^c$ to have positive  $||G_\Delta(z)||_m>0$ in the range $ 1-z>\epsilon_0$ for any $\Delta>\Delta_{\textrm{TP}}^c$. The numerical precision is $10^{-8888}$.  
\label{DetGm}
}
 \begin{center}
\begin{tabular}{cccccccc}
\hline\hline\\[-.5em] $m$ &
			  ~3~ & ~5~ & ~7~ & ~9 ~&~  11 ~ ~&~  13 ~ ~&~  15 ~
 \\[.5em]\hline\\ [-.5em]
      $\epsilon_0$ 
			& ~ $10^{-5000}$~ & $10^{-3000}$~& $10^{-2500}$~& $10^{-1200}$ ~&  $10^{-1000}$ ~ &  $10^{-800}$ ~ &  $10^{-600}$~ 
 \\[.5em]\hline\\ [-.5em]  $\Delta_{\textrm{TP}}^c\approx$ 
			& ~ 0.1627~ & 0.1319~& 0.108~& 0.0885 ~&  0.076 ~ &  0.066 ~ &  0.059~  \\[.5em]\hline\hline 
		\end{tabular}
 \end{center}
\end{table}

\begin{table}
\caption{Critical  value $\epsilon^c$ for a given $\Delta_{\textrm{TP}} $ with which $||G_\Delta(z)||_{m=3}<0   ~\textrm{for}~  \epsilon<\epsilon^c$.  
\label{DetGm3}
}
 \begin{center}
\begin{tabular}{cccccccccc}
\hline\hline\\ [-.5em]  $\Delta_{\textrm{TP}}$ 
			& 0.01 & 0.1 & ~ 0.14~ & 0.15~& 0.16~& 0.162 ~&  0.1626 ~ &  0.16264 ~ &  0.1627~  \\[.5em]\hline\\ [-0.5em]
      $\epsilon^c\approx$ 
			&~ $10^{-2}$ &~ $10^{-4}$ & ~ $10^{-9}$~ & $10^{-15}$~& $10^{-67}$~& $10^{-245}$ ~&  $10^{-1357}$ ~ &  $10^{-1949}$ ~ &  $10^{-5647}$~ 
 \\[.5em]\hline\hline 
		\end{tabular}
 \end{center}
\end{table}

In Table \ref{DetGm3} we show the range of $\epsilon^c$ below which the 
 third order determinant $||G_\Delta(z)||_{m=3}$ becomes negative for a given $\Delta_{\textrm{TP}}$. 
Near the threshold value $\Delta_{\textrm{TP}}^*\approx 0.1627$ the range of variable $|1-z|=\epsilon<\epsilon^c$ for negative determinant becomes extremely small! At $\Delta_{\textrm{TP}}= 0.1627$  the positivity of the determinant $||G_\Delta(z)||_{m=3}$ is only violated by a tiny factor at the order 
\begin{equation}
   ||G_\Delta(z)||_{m=3}\approx  \left. -1.5939\times 10^{-5654}\right.|_{\Delta_{\textrm{TP}=0.1627}}.
\end{equation}
The total positivity of 1D conformal block $G_\Delta(z)$ is so sophisticated that for a normal  parameter   with four effective digits at the order $10^{-1}$, it is merely violated by a negative $3\times 3$ determinant at the order $10^{-5654}$!  
Such a ``hierarchy"  naturally arising from the total positivity of the 1D conformal block has a span of $5653$ orders, drastically larger than the famous hierarchy problem between the electroweak scale ($10^2$ GeV) and the Planck scale ($10^{19}$ GeV)! The hidden mathematical structure is even more astonishing than the seemingly unnatural parameters. It inspires a question that could the hierarchy problem in particle physics be related to certain positive structure in quantum field theories?

Behavior of $||G_\Delta(z)||_{m=3}$ near the threshold value $\Delta_{\textrm{TP}}^c$ can be studied analytically. Let us consider the small $\epsilon$ expansion of the conformal block $G_\Delta(1-\epsilon)$ 
\begin{align}
    \frac{\Gamma (\Delta )^2}{\Gamma (2\Delta )} G_\Delta(1-\epsilon) &=  \\
    & \hspace{-2.3cm} -2 \psi ^{(0)}(\Delta )-\log (\epsilon )-2 \gamma-\epsilon\,\Delta(\Delta -1)   \left(2 \left(\psi ^{(0)}(\Delta )+\gamma \right)+\log (\epsilon )-2\right)+O(\epsilon^2),  \nn
\end{align}
where $\gamma$ is the Euler constant and $\psi^{(0)}$ is the zeroth order Polygama function.
With the data set (\ref{Dataset}), the sign of the determinant $||G_\Delta(z)||_{3}$ is given by
\begin{equation}
    ||G_\Delta(z)||_{3}\propto\epsilon \left(\mathbf{P}(\Delta_{\textrm{TP}})-\mathbf{Q}(\Delta_{\textrm{TP}})\log(\epsilon)\right)+O(\epsilon^2), \label{DetCBepsilon}
\end{equation}
where
\begin{align}
    \mathbf{Q}(x)=& 2 x \log \left(4/3\right)\times \nn \\
    &\left(\psi ^{(0)}(x)-2 \psi ^{(0)}(2 x)+\psi ^{(0)}(3 x)-x  \left(5 \psi ^{(0)}(x)-8 \psi ^{(0)}(2 x)+3 \psi ^{(0)}(3 x)\right)\right).
\end{align}
In the limit $\epsilon\rightarrow0, ~\log(\epsilon)\rightarrow -\infty$, the sign of  $||G_\Delta(z)||_{3}$ is determined by the factor $\mathbf{Q}(\Delta_{\textrm{TP}})$ in (\ref{DetCBepsilon}). It is straightforward to check that the function $\mathbf{Q}(x)$ has a unique positive root at $x^c\approx0.1627316$ and is always positive for $x>x^c$. 
Therefore in the small $\epsilon$ limit, the determinant $||G_\Delta(z)||_{3}$ with arguments (\ref{Dataset}) is positive for $\Delta>\Delta_{\textrm{TP}}^c\approx0.1627316$,  beautifully consistent with our high precision numerical results in Tables \ref{DetGm} and \ref{DetGm3}.

Now let us consider the determinant $||G_\Delta(z)||_{3}$ with more general $\Delta_i$ and $z_j$:
\begin{equation}
    ~~\Delta_i=\Delta_{\textrm{TP}}\, w_i, ~~~ z_j=1+\epsilon\,y_j,   ~~~~ 1=w_1<w_2<w_3,~~ y_1<y_2<y_3=-1 . \label{Dataset1}
\end{equation}
In the small $\epsilon$ limit the determinant $||G_\Delta(z)||_{3}$ has similar formula as (\ref{DetCBepsilon}).
In particular its dominating part is also given by the term proportional to $\epsilon \log(\epsilon)$ and the function $\mathbf{Q}$ is modified to
\begin{align}
  \mathbf{Q}(x)=& 2 x \left(y_1  \log \left(y_2/y_3\right) +y_2  \log \left(y_3/y_1\right) +y_3  \log \left(y_1/y_2\right) \right)\times \nn\\
  &   \left(\left(w_3-w_2\right) \psi ^{(0)}(x)-\left(w_3-1\right) \psi ^{(0)}\left(x w_2\right) +\left(w_2-1\right) \psi ^{(0)}\left(x w_3\right)+\right.  \label{Sgeneral}\\
  &\left.~~x \left( \left(w_2^2-w_3^2\right) \psi ^{(0)}(x) -(w_2^2-1 ) \psi ^{(0)}\left(x w_3\right) +\left(w_3^2-1\right) \psi ^{(0)}\left(x w_2\right)\right)\right), \nn
\end{align}
in which the $y_i$-dependent term is always positive. Note the non-even distribution of the variables $z_j$ has trivial effect on the sign of $\mathbf{Q}(x)$, as in the small $\epsilon\rightarrow 0$ limit, the factors $y_i$ are decoupled from the dominating term $\propto \log(\epsilon)$.
The equation $\mathbf{Q}(x)=0$ can be solved numerically or using small $w_2-1,~ w_3-1$ expansion. The equation has a unique solution $x^c$ above  which $\mathbf{Q}(x)>0$. The root $x^c$ depends on $w_2, w_3$ and it reaches the maximum value in the limit $w_2, w_3\rightarrow 1$, which is given by the equation
\begin{equation}
    2 \psi ^{(1)}(x)+(1-2 x) \psi ^{(2)}(x)=0.
\end{equation}
The solution to this equation is $x^*\approx 0.32315626$. It gives the maximum value of $\Delta_{\textrm{TP}}$ to have a negative $||G_\Delta(z)||_{3}$ for general data set $\{\Delta_i, z_j\}$.

Let us summarize what we have obtained so far. From the evenly distributed data $\{\Delta_i, z_j\}$, the numerical results suggest $||G_\Delta(z)||_{m}$ can have negative values below a threshold value $\Delta_{\textrm{TP}}<\Delta_{\textrm{TP}}^c$ for small $1-z$. The $\Delta_{\textrm{TP}}^c$ decreases with larger $m$ and obtains its maximum value at $m=3$:
\begin{equation}
 \left.\Delta_{\textrm{TP}}^c\right|_{m=3}>\Delta_{\textrm{TP}}^c|_{m>3}. \label{DetlaTPIeq}
\end{equation}
From a careful analysis for $||G_\Delta(z)||_{m}$ with general   $\{\Tilde{\Delta}_{i}, \Tilde{z}_{j}\}$ at $m=3$, we find the $\Delta_{\textrm{TP}}^c$ reaches its maximum value  near the cusp $0<w_2-1<w_3-1\ll1$. If the inequality (\ref{DetlaTPIeq}) is also true for non-evenly distributed data $\{\Tilde{\Delta}_{i}, \Tilde{z}_{j}\}\bigcup\{\Delta_{i>3}, z_{j>3}\}$, then our results suggest the solution $x^*$ is optimal and the 1D $\SL$ conformal block is totally positive for any $\Delta>\Delta_{\textrm{TP}}^*\approx 0.32315626$.

\subsubsection*{Total positivity of the linear functional action on $G_\Delta(z)$} 
In Section \ref{sec4} we will study the functional $\alpha'_i$ whose action  is given by
\begin{equation}
   S(\Delta,i)\equiv\alpha'_i[G_\Delta]= \int_0^1 x^{i+a} \,G_\Delta(x)dx. \label{Action1}
\end{equation}
and it is important to know the total positivity of the function $S(\Delta,i)$.

With sufficiently large $\Delta$, the total positivity of the function $S(\Delta, i)$ can be proved using the basic composition formula (\ref{BCf}). Since the function $z^{i+a}$ and $G_\Delta(z)$ with large $\Delta$ are totally positive, their convolution is also totally positive. 
Note the total positivity of $G_\Delta(z)$ is a sufficient but not necessary condition for $S(\Delta, i)$ being totally positive, and it could be totally positive  with small $\Delta_i$ though this is not the case for $G_\Delta(z)$. The integration (\ref{Action1}) can be evaluated using series expansion of $G_\Delta(z)$:
\begin{equation}
    S(\Delta,i)=  \sum\limits_{k=0}^\infty \frac{1}{k!} \frac{(\Delta)_k^2}{(2\Delta)_k}\frac{1}{k+i+a+1}. \label{TPGaction}
\end{equation}
In the above formula, the function $\frac{1}{k+i+a+1}$ is the modified Cauchy's matrix which is totally positive, see Appendix 
\ref{TPEx2}. For another relevant factor $\frac{(\Delta)_k^2}{(2\Delta)_k}$, we have provided promising evidence for its total positivity before. Therefore the linear functional action $S(\Delta,i)$ is also expected to be totally positive for $\Delta\geqslant0,~ i\in \mathbb{N}$.

\section{Analytical functionals for the 1D $O(N)$ vector bootstrap bound}\label{sec4}
In this section we construct the analytical functionals for the  1D large $N$ bootstrap bound with $\Delta\in(0,\Delta_c/2)$, which is saturated by the the generalized free field theory with spectrum $\Delta_n=2\Delta_\phi+n, ~n\in \mathbb{N}$ in the $T$ sector. By constructing the analytical functionals for this simple while representative bootstrap problem, we want to study the critical question in conformal bootstrap: what is the mathematical structure responsible for the nontrivial bootstrap constraints? 
To construct the analytical functionals, we utilize the functional basis dual to the spectrum of generalized free field theories \cite{Mazac:2019shk}, for which we review in the first part of this section.

\subsection{Analytical functional basis}
In Section \ref{sec2}, the linear functionals are constructed based on the derivatives of variable $z$ at the crossing symmetric point $z=\frac{1}{2}$: $\alpha=\sum_{i\leqslant \Lambda} c_i\;\partial^i_z\cdot|_{z=\frac{1}{2}}$. These functionals are convenient for numerical computations. Nevertheless, due to the singularities at $z=0, 1$ of the conformal block $G_\Delta(z)$,  the series expansion of $G_\Delta(z)$ only converges in the range  $|z-\frac{1}{2}|<\frac{1}{2}$.
To construct functionals more effectively,  it needs new basis which contains information of the singularities of $G_\Delta(z)$, namely the analytical functional basis \cite{Mazac:2016qev,Mazac:2018mdx,Mazac:2018ycv}.  

The  analytical functional basis is dual to the function basis in terms of which the conformal correlation functions can be expanded.  The function basis can be provided by the s- and t-channel conformal blocks
\begin{equation}
    G_n^s\equiv z^{-2\Delta_\phi}\;G_{\Delta_n}(z), ~~~~~G_n^t\equiv (1-z)^{-2\Delta_\phi}\;G_{\Delta_n}(1-z),
\end{equation}
and their derivatives, associated with the spectrum of generalized free field theories, e.g., the generalized free boson $\Delta_n=2\Delta_\phi+2n$  or fermion $\Delta_n=2\Delta_\phi+2n+1$ \cite{Mazac:2018mdx,Mazac:2018ycv}.
In this work, inspired by the extremal functional spectrum in Fig. \ref{EFM}, we adopt a different function basis for the conformal correlation function, which is given by the conformal blocks $ G_{n}^s,~G_{n}^t $ without their derivatives, associated with the spectrum $\Delta_n=2\Delta_\phi+n,~n\in \mathbb{N}$. 
Consider a correlation function $\cG(z)$ which is superbounded in the u-channel Regge limit $|z|\rightarrow \infty$:\footnote{Here the correlation function $\cG(z)$ is the correlation function $\cG(z)$ in  (\ref{corrf}) dressed with a factor $z^{-2\Delta_\phi}$.}
\begin{equation}
    |\cG(z)|<|z|^{-\epsilon}
\end{equation}
with $\epsilon>0$, it admits a unique expansion in terms of the above function basis 
\begin{equation}
    \cG(z)=\sum\limits_{n=0}^\infty \; \lambda_n^s \; G_n^s + \sum\limits_{n=0}^\infty \;  \lambda_n^t \; G_n^t\equiv \cG^s(z)+\cG^t(z). \label{Gexp}
\end{equation}
The basis $G_n^s$ is holomorphic away from $z\in [1,+\infty)$, so is $\cG^s(z)$. Likewise, the function $\cG^t(z)$ is holomorphic away from $z\in (-\infty, 0]$.
The functional basis $\alpha_m^{s,t}$ dual to the above function basis satisfies 
\begin{align}
    \alpha_m^s\cdot G_n^s &=\delta_{mn},  ~~~~~\alpha_m^s\cdot G_n^t =0, 
 \label{act1}\\
\alpha_m^t\cdot G_n^t &=\delta_{mn}, ~~~~~ \alpha_m^t\cdot G_n^s =0,\label{act2}
\end{align}
 based on which the coefficients $\lambda_n^{s/t}$ in (\ref{Gexp}) can be extracted from the Regge superbounded conformal correlator 
 \begin{equation}
     \lambda_n^s=\alpha_n^s\cdot\cG,~~~\lambda_n^t=\alpha_n^t\cdot\cG
 \end{equation}
and the expansion (\ref{Gexp}) can be formally rewritten as
\begin{equation}
\cG(z)=\cG^s(z)+\cG^t(z)=\sum\limits_{n=0}^\infty \,(\alpha_n^s\cdot \cG)\; G_n^s+(\alpha_n^t\cdot \cG)\;G_n^t . \label{Gexp1}
\end{equation}
Above formula has close relation with the dispersion relation  of conformal correlation functions \cite{Mazac:2019shk,Bissi:2022mrs}. Here we sketch the main idea. Consider the Cauchy's integral formula for the conformal correlation function $\cG(z)$:
\begin{equation}
    \cG(z)=\oint\frac{dw}{2\pi i}\;\frac{1}{w-z}\;\cG(w),
\end{equation}
in which the contour encircles $w=z$ but does not contact the branch cuts $(-\infty,0]$ and $[1,+\infty)$.
The contour can be deformed into contours wrapping the two branch cuts, denoted $C_{\mp}$ and the arcs at infinity. For the Regge superbounded correlation functions which satisfy $\cG(w)=O(|w|^{-\epsilon})$ in the Regge limit $|w|\rightarrow \infty$, contributions from infinity vanish and the integral of $\cG(z)$ consists of two parts 
\begin{equation}
    \cG(z)=-\int_{C_-}\frac{dw}{2\pi i}\;\frac{1}{w-z}\;\cG(w)+ \int_{C_+}\frac{dw}{2\pi i}\;\frac{1}{w-z}\;\cG(w)\equiv \cG^t(z)+\cG^s(z), \label{Intparts}
\end{equation}
in which the $\cG^t(z)$ and $\cG^s(z)$ are  holomorphic away from $z\in(-\infty, 0]$ and $z\in[(1,+\infty)$, respectively. The holomorphicity of the two terms in  (\ref{Intparts}) suggests they can be decomposed into the function basis of $G_n^t$ and $G_n^s$, as in (\ref{Gexp}).
Such decomposition can be alternatively fulfilled with the expansion of the integral kernel $\frac{1}{w-z}$ 
\begin{equation}
    \frac{1}{w-z}=\sum\limits_{n=0}^\infty \Theta_n(w) \; z^{-2\Delta_\phi} G_{\Delta_n}(z)
\end{equation}
for integral along the contour $C_+$ and 
 \begin{equation}
    \frac{1}{w-z}=-\sum\limits_{n=0}^\infty \Theta_n(1-w) \;(1-z)^{-2\Delta_\phi}G_{\Delta_n}(1-z)
\end{equation}
for integral along the contour $C_-$,
in which  
\begin{equation}
    \Theta_n(w)=\frac{(-1)^n(2\Delta_\phi)_n^2}{n!(4\Delta_\phi+n-1)_n}\frac{1}{w} \;_3F_2(1,-n,4\Delta_\phi+n-1;2\Delta_\phi,2\Delta_\phi;\frac{1}{w}). \label{Theta}
\end{equation}
The integrals in (\ref{Intparts}) turn into
\begin{align}
    \cG^s(z)=\int_{C_+}\frac{dw}{2\pi i}\;\frac{1}{w-z}\;\cG(w)&=\sum\limits_{n=0}^\infty \left(\int_{C_+}\frac{dw}{2\pi i}\Theta_n(w)\;\cG(w) \right)G_n^s \;, \\
    \cG^t(z) =-\int_{C_-}\frac{dw}{2\pi i}\;\frac{1}{w-z}\;\cG(w)&=\sum\limits_{n=0}^\infty \left(\int_{C_-}\frac{dw}{2\pi i} \Theta_n(1-w)\;\cG(w) \right)G_n^t\;.
\end{align}
Comparing with the expansion (\ref{Gexp1}), it gives explicit formulas for the actions of the functional basis
\begin{align}
    \alpha_n^s\cdot \cG &=\int_{C_+}\frac{dz}{2\pi i} \Theta_n(z)\;\cG(z) , \label{Ints}\\
    \alpha_n^t\cdot \cG &=\int_{C_-}\frac{dz}{2\pi i} \Theta_n(1-z)\;\cG(z).\label{Intt}
\end{align}
By deforming the contours it is clear that the actions $\alpha_n^s\cdot G_n^t=\alpha_n^t\cdot G_n^s=0$.

There are constraints the analytical functionals need to satisfy  \cite{Qiao:2017lkv, Mazac:2018mdx, Mazac:2018ycv}.
Here $\Theta_n(z)\sim O(|z|^{-1})$ in the Regge limit $|z|\rightarrow \infty$, therefore the integrals (\ref{Ints},\ref{Intt}) only converge for the functions $\cG(z)\sim O(|z|^{-\epsilon})$ in this limit. 
The most general conformal correlation functions have Regge limit $\cG(z)\rightarrow |z|^0$ for which the integrals  (\ref{Ints},\ref{Intt}) do not converge. In this case one can use the subtracted functionals \cite{Mazac:2019shk}
\begin{equation}
    \bar{\alpha}_n^r=\alpha_n^r-\frac{(-1)^n(2\Delta_\phi)_n^2}{n!(4\Delta_\phi+n-1)_n}\alpha_0^r,  ~~~~ r=s,t, \label{alphabar}
\end{equation}
which correspond to new integral kernels with Regge behavior $\bar{\Theta}_n(z)\sim O(|z|^{-2})$.

\subsubsection*{Actions of the functional basis on conformal blocks }
The dual relations (\ref{act1},\ref{act2}) show the actions of functional basis on the conformal blocks with $\Delta=2\Delta_\phi+n$. For the actions on conformal blocks with general $\Delta$'s, we need to evaluate the integrals (\ref{Ints}) with $\cG=z^{-2\Delta_\phi}G_\Delta(z)$ and $\cG=(1-z)^{-2\Delta_\phi}G_\Delta(1-z)$
\begin{align}
    \alpha_n^s\cdot \left(z^{-2\Delta_\phi}G_\Delta(z)\right)\equiv \mathbf{S}(\Delta,n) &=\int_1^\infty \frac{dz}{2\pi i}\textrm{Disc}[\Theta_n(z)z^{-2\Delta_\phi}G_\Delta(z)], \label{alphaGs}\\
    \alpha_n^s\cdot \left((1-z)^{-2\Delta_\phi}G_\Delta(1-z)\right)\equiv \mathbf{T}(\Delta,n) &=\int_1^\infty \frac{dz}{2\pi i}\textrm{Disc}[\Theta_n(z)(1-z)^{-2\Delta_\phi}G_\Delta(1-z)].~~ \label{alphaGt}
\end{align}
The function $\Theta_n(z)$ is regular along $z\in [1,+\infty)$, while the conformal blocks acquire discontinuities between the two sides of the branch cut  $[1,+\infty)$
\begin{align}
   \frac{1}{2\pi i} \textrm{Disc}[z^{-2\Delta_\phi}G_\Delta(z)]&= \frac{\Gamma(2\Delta)}{\Gamma(\Delta)^2} \; z^{-2\Delta_\phi-\Delta+1} \,_2F_1(1-\Delta,1-\Delta,1,1-z), \nn\\
   \frac{1}{2\pi i} \textrm{Disc}[(1-z)^{-2\Delta_\phi}G_\Delta(1-z)]&=-\frac{\sin(\pi(\Delta-2\Delta_\phi))}{\pi}  (z-1)^{\Delta-2\Delta_\phi}\,_2F_1(\Delta,\Delta,2\Delta,1-z).  \nn
\end{align}
Applying above two formulas in (\ref{alphaGs}) and (\ref{alphaGt}) it gives 
\begin{align}
    \mathbf{S}(\Delta,n)=&\frac{(-1)^n \Gamma(2\Delta)\Gamma (n+2 \Delta_\phi )^2  }{\Gamma (n+1)\Gamma(\Delta)^2 \Gamma (-\Delta +2 \Delta_\phi +1) \Gamma (\Delta +2 \Delta_\phi ) (n+4 \Delta_\phi -1)_n} \nn\\
    &\times\,_3F_2(1,-n,n+4 \Delta_\phi -1;-\Delta +2 \Delta_\phi +1,\Delta +2 \Delta \phi ;1), \label{alphas1}
\end{align}
and 
\begin{align}
    \mathbf{T}(\Delta,n)=\frac{\sin (\pi  (\Delta -2 \Delta_\phi ))}{\pi }   \sum _{i=0}^n &\frac{ (-1)^{i+n+1}\Gamma (2 \Delta )  \Gamma (\Delta -2 \Delta _\phi +1) \Gamma (2 \Delta_ \phi+i ) (2 \Delta _\phi +i)_{n-i}^2}{\Gamma (n-i+1)  (4 \Delta _\phi +n+i-1)_{n-i}} \nn\\
    & ~\times\, _3\tilde{F}_2(\Delta ,\Delta ,\Delta -2 \Delta_ \phi +1;2 \Delta ,\Delta +i+1;1),\label{alphat1}
\end{align}
where $\, _3\tilde{F}_2$ is the regularized generalized hypergeometric function.
For $\Delta=2\Delta_\phi+m,~ m\in \mathbb{N}$, above formulas agree with the dual relation (\ref{act1}). Actions of the linear functionals $\alpha_n^t$ on the s- and t-channel conformal blocks can be obtained from (\ref{alphas1}) and (\ref{alphat1}) through $z\leftrightarrow 1-z$ transformation.

For $\Delta_\phi=\frac{1}{2}$ the kernel $\Theta_n^s(z)$ in (\ref{act1}) is drastically simplified
\begin{equation}
    \Theta_n(z)= \frac{1}{z}\frac{ (-1)^n \Gamma (n+1)^2  }{\Gamma (2 n+1)}\, _2F_1\left(-n,n+1;1;\frac{1}{z}\right). \label{Thetahalf}
\end{equation}
Its actions on conformal blocks are reduced to
\begin{equation}
    \mathbf{S}(\Delta,n)|_{\Delta_\phi=\frac{1}{2}}=\frac{ \Gamma (2 \Delta ) \Gamma (n+1)^2 \sin (\pi  (\Delta -n-1))}{\pi  \Gamma (\Delta )^2 (\Delta -n-1) (\Delta +n) \Gamma (2 n+1)},\label{alphas2}
\end{equation}
and 
\begin{align}
    \mathbf{T}(\Delta,n)|_{\Delta_\phi=\frac{1}{2}}=\,&
    \frac{(-1)^n (n!)^2 \,\Gamma (2 \Delta )\,\Gamma (\Delta )^2  }{\Gamma (2 n+1)}\frac{\sin(\pi\Delta)}{\pi} \nn\\
    &\times\, _4\tilde{F}_3(\Delta ,\Delta ,\Delta ,\Delta ;2 \Delta ,\Delta -n,\Delta +n+1;1).\label{alphat2}
\end{align}
Above formulas provide necessary ingredients to construct analytical functionals. 

\subsection{Analytical functionals for Regge superbounded conformal correlator}
In Section \ref{sec2} we have shown that in the range $\Delta<\Delta_c$, the $O(\infty)$ vector bootstrap bound on the scaling dimension of the lowest operator in the $T$ sector $\Delta_T^*$ is determined by the crossing equation
\beq 
\sum_{\cO\in S}\lambda_{\cO}^2 \,z^{-2\Delta_\phi}\,G_{\Delta}(z) - \sum_{\cO \in T}\lambda_{\cO}^2
\,(1-z)^{-2\Delta_\phi}\, G_{\Delta}(1-z)=0,
\label{simpceq2}
\eeq  
which is saturated by $\Delta_T^*=2\Delta_\phi$ with extremal spectrum $2\Delta_\phi+n, ~n\in\mathbb{N}$.
Consequently the extremal functional $\alpha^*$ should satisfy following positive conditions
\begin{align}
    \alpha^*\cdot \left(z^{-2\Delta_\phi}G_\Delta(z)\right) &=0, ~~~\textrm{for}~~ \Delta=0, \label{cd1} \\
    \alpha^*\cdot\left(z^{-2\Delta_\phi}G_\Delta(z)\right)&>0, ~~~\forall ~\Delta>0, \label{cd2}\\
   -\alpha^*\cdot \left((1-z)^{-2\Delta_\phi}G_\Delta(1-z)\right)&=0^1, ~~\textrm{for}~~ \Delta=2\Delta_\phi, \label{cd3}\\
   -\alpha^*\cdot \left((1-z)^{-2\Delta_\phi}G_\Delta(1-z)\right)&=0^2, ~~\textrm{for}~~ \Delta=2\Delta_\phi+n, n\in \mathbb{N}^+, \label{cd4}\\
    -\alpha^*\cdot \left((1-z)^{-2\Delta_\phi}G_\Delta(1-z)\right)&>0, ~~~\forall ~\Delta>2\Delta_\phi ~\& ~\Delta\neq 2\Delta_\phi+n, n\in \mathbb{N}^+,\label{cd5}
\end{align}
in which the notation $0^i$ refers to the zeros of order $i$. In (\ref{cd4}) the zeros $0^{2i}$ with $i>1$    also satisfy the positive condition, however, we will show that there are only second order zeros in (\ref{cd4}). 
 
For the Regge superbounded conformal correlators, e.g., correlation functions (\ref{GFF1}-\ref{GFF3}) with $\lambda=0$, the bootstrap functional $\alpha$ can be expanded in terms of the functional basis 
\begin{equation}
    \alpha=\sum\limits_{n=0}^\infty c_n \alpha_n^s+d_n \alpha_n^t.
\end{equation}
Considering the actions of the functional basis on $G_n^{s,t}$ (\ref{act1},\ref{act2}), the positive condition (\ref{cd2}) of the  extremal functional $\alpha^*$ requires 
\begin{equation}
    c_n>0 ~~~ \forall n\in \mathbb{N}. \label{postivec}
\end{equation}
Moreover, the conditions (\ref{cd3},\ref{cd4}) suggest
\begin{equation}
    d_n=0 ~~~\forall n\in\mathbb{N}.
\end{equation}
The positive coefficients $c_n$ in $\alpha^*=\sum\limits_{n=0}^\infty c_n \alpha_n^s$ should be arranged so that the extra positive conditions can be satisfied.

It is easy to verify that the positive condition (\ref{cd1}) is  satisfied by the functional $\alpha^*=\sum\limits_{n=0}^\infty c_n \alpha_n^s$. For each $\alpha_n^s$, its action is 
\begin{equation}
    \alpha_n^s\cdot z^{-2\Delta_\phi} =\int_{C_+}\frac{dz}{2\pi i} \Theta_n(z)\;z^{-2\Delta_\phi},
\end{equation}
in which the integrand has a pole at $z=0$ and is holomorphic for $z\in[1,+\infty)$ enclosed by the contour $C_+$. Therefore the action vanishes, as required by (\ref{cd1}).

The critical constraints to solve $c_n$ are from (\ref{cd3},\ref{cd4}). 
In the  action (\ref{alphat1}) of functional basis
$\alpha_n^s$, the factor $\sin(\pi(\Delta-2\Delta_\phi))$ generates single zeros at $\Delta=2\Delta_\phi+n$ with $n\in \mathbb{N}$. To further form double zeros for $n\in \mathbb{N}^+$, the coefficients $c_n$ should satisfy 
\begin{equation}
    \sum_{i=0}^\infty\mathcal{T}(2\Delta_\phi+n,i)\cdot \Tilde{c}_i=\delta_{n0}, ~~~\forall n\in\mathbb{N} \label{eqc}
\end{equation}
 in which  the coefficients $\Tilde{c}_i$ is
\begin{equation}
   \Tilde{c}_i\equiv c_i \,(-1)^i\;\frac{ \Gamma (i+1)^2}{\Gamma (2 i+1)}.
\end{equation}
 and
 $\mathcal{T}$ is the stripped action
 \begin{equation}
     \mathbf{T}(\Delta,i)=  (-1)^{i+1} \;\frac{\sin (\pi  (\Delta -2 \Delta_\phi ))}{\pi } \;\mathcal{T}(\Delta,i)\;\frac{ \Gamma (i+1)^2}{\Gamma (2 i+1)}.
 \end{equation}
One may expect the coefficients $c_n$ can be obtained by solving the whole infinite equation group (\ref{eqc}), like the remarkable work \cite{Mazac:2016qev}. However, the solutions to the whole equation group (\ref{eqc}) lead to a trivial functional. We demonstrate this point using an example with $\Delta_\phi=\frac{1}{2}$. 

\subsubsection{Solution of the infinite equation group}
Solution to the linear equation group (\ref{eqc}) is given by the inverse of the infinite matrix $\mathcal{T}(2\Delta_\phi+n,i)$. 
For general $\Delta_\phi$ the matrix  $\mathcal{T}(2\Delta_\phi+n,i)$ is quite complicated and hard to solve directly. The formulas are notably simplified with $\Delta_\phi=\frac{1}{2}$, see (\ref{Thetahalf}-\ref{alphat2}).
In this case it is convenient to take the variable transformation $x=1-z^{-1}$ \cite{Mazac:2016qev}.
The kernel $\Theta_n(z)$ degenerates to the Legendre polynomials $P_{n}(2x-1)$ and the stripped action $\mathcal{T}(\Delta,n)$  becomes
\begin{equation}
    \mathcal{T}(\Delta,n)=\int_0^1 dx \;x^{\Delta-1}\,_2F_1(\Delta,\Delta,2\Delta,x) P_{n}(2x-1). \label{Tforhalf}
\end{equation}
Since the Legendre polynomials $P_n(2x-1)$ are orthogonal under the inner product
\begin{equation}
    \int_0^1dx\; P_n(2x-1)P_m(2x-1)=\frac{1}{2n+1}\delta_{n,m},
\end{equation}
the matrix  $\mathcal{T}(m,n)$ can be interpreted as the coefficients of the following expansion
\begin{equation}
    x^{m-1}\,_2F_1(m,m,2m,x)=\sum_{n=0}^\infty (2n+1) \mathcal{T}(m,n)P_n(2x-1).
\end{equation}
Thus the inverse matrix $\mathcal{T}^{-1}$ is given by the inverse expansion 
\begin{equation}
   P_n(2x-1)=\frac{1}{2n+1}\sum_{m=1}^\infty \mathcal{T}^{-1}(n,m)~ x^{m-1}\,_2F_1(m,m,2m,x). \label{P2F1}
\end{equation}
Let us compare the constant term on both sides for each $n$.
The Legendre polynomial satisfies $P_n(2x-1)|_{x=0}=(-1)^n$. While the function $x^{m-1}\,_2F_1(m,m,2m,x)$ is equal to 1 at $x=0$ for $m=1$ and
vanishes at $x=0$ for $m>1$.
Therefore the elements $\mathcal{T}^{-1}(n,1)$, as well as the coefficients $\Tilde{c}_n$ in (\ref{eqc}) can be solved from (\ref{P2F1}):
\begin{equation}
    \mathcal{T}^{-1}(n,1)=\Tilde{c}_n=(-1)^n(2n+1), \label{largecn}
\end{equation}
which gives
\begin{equation}
    c_n=\frac{\Gamma(2n+2)}{\Gamma(n+1)^2}.
\end{equation}
A comment is that all the $c_n$'s are positive, as needed to satisfy the positive condition (\ref{cd2}). 

The whole extremal functional 
$\alpha^*$ is given by 
\begin{equation}
    \alpha^*\cdot \cG(z)=\int_{C_+}\frac{dz}{2\pi i}\Theta^*(z)\cG(z)  
\end{equation}
with a kernel
\begin{equation}
    \Theta^*(z)=\sum_{n=0}^\infty  c_n \Theta_n(z)\propto \sum_{n=0}^\infty \left.(-1)^n (2n+1) P_n(2x-1)\right|_{x=\frac{z-1}{z}}\equiv \bar{\Theta}^*(x).
\end{equation}
Using the generating function of Legendre polynomial
\begin{equation}
    \frac{1}{\sqrt{t^2-2 t x+1}}=\sum_{n=0}^\infty P_n(x)\;t^n,
\end{equation}
one can show
\begin{equation}
    \bar{\Theta}^*(x)= \left.\frac{1-t^2}{\left(4 t x+(1-t)^2\right)^{3/2}}\right|_{t\rightarrow 1}.
\end{equation}
In the limit $t\rightarrow 1$, we have
\begin{equation}
    \bar{\Theta}^*(x)|_{x\neq0}=0.
\end{equation}
While with $x=0$ the function $\bar{\Theta}^*(x)$ has a pole at  $t=1$. Such ``extremal" kernel behaves like a Dirac $\delta$-function $\delta(x)$.
The action of functional $\alpha^*$ on the t-channel conformal block $ -(1-z)^{-2\Delta_\phi}G_\Delta(1-z) $ with $\Delta_\phi=\frac{1}{2}$ becomes
\begin{align}
     -\alpha^*&\cdot \left((1-z)^{-1} G_\Delta(1-z)\right)= 
 \nn\\
 &
 \hspace{1.6cm}\frac{\sin(\pi(\Delta-1))}{\pi}\int_0^1 dx \left.\frac{1-t^2}{\left(4 t x+(1-t)^2\right)^{3/2}} \;x^{\Delta-1} \,_2F_1(\Delta,\Delta,2\Delta,x)\right|_{t\rightarrow1}.
\end{align}
Due to the pole at $x=0$ with $t=1$, above integral only gives a nonzero value for $\Delta=2\Delta_\phi=1$, while vanishes for $\Delta>1$. Together with the factor $\sin(\pi(\Delta-1))$, the whole action vanishes for all $\Delta\geqslant 1$.
On the other hand, its action on the s-channel conformal block
\begin{equation}
     \alpha^*\cdot \left(z^{-1}G_\Delta(z)\right)=\frac{\Gamma(2\Delta)}{\Gamma(\Delta)^2} \int_0^1 dx \left.\frac{1-t^2}{\left(4 t x+(1-t)^2\right)^{3/2}} \;  \,_2F_1(\Delta,1-\Delta,1,x)\right|_{t\rightarrow1}
\end{equation}
vanishes at $\Delta=0$ and is always positive for $\Delta>0$.

To summarize, the functional constructed from the whole infinity set of equations (\ref{eqc}) is actually trivial due to its vanishing action on the t-channel conformal block, or the $O(N)$ $T$ sector in (\ref{simpceq2}).
We expect this is the case for general $\Delta_\phi$.

\subsubsection{Solutions of the finite subset of equation group}
Although the inverse of the whole equation group (\ref{eqc}) leads to a degenerated functional, surprisingly the inverse of the finite subset of the equation group  can produce functionals which satisfy all the positive conditions (\ref{cd1}-\ref{cd5}) within a range $\Delta\leqslant \Lambda_N$. This allows us to construct a series of functionals with arbitrarily high  $\Lambda_N$!

Instead of constructing a functional whose action on t-channel conformal block has double zeros at $\Delta=2\Delta_\phi+n$
for all $n\in \mathbb{N}^+$, we would like to relax the restriction on the double zeros. Specifically, we consider the functional 
\begin{equation}
    \alpha'_N=\sum_{n=0}^{N}c_n\alpha_n^s,
\end{equation}
whose action  has a single zero at $\Delta=2\Delta_\phi$ and double zeros at 
$\Delta=2\Delta_\phi+n$
for each integer $0<n\leqslant N$. This amounts to the following constraints
\begin{equation}
    \sum_{i=0}^N\mathcal{T}(2\Delta_\phi+n,i)\cdot \tilde{c}_i =\delta_{n0}, 
  ~~~~~0\leqslant n\leqslant N.  \label{eqc1}
\end{equation}

\begin{figure}
\includegraphics[width=0.48\linewidth]{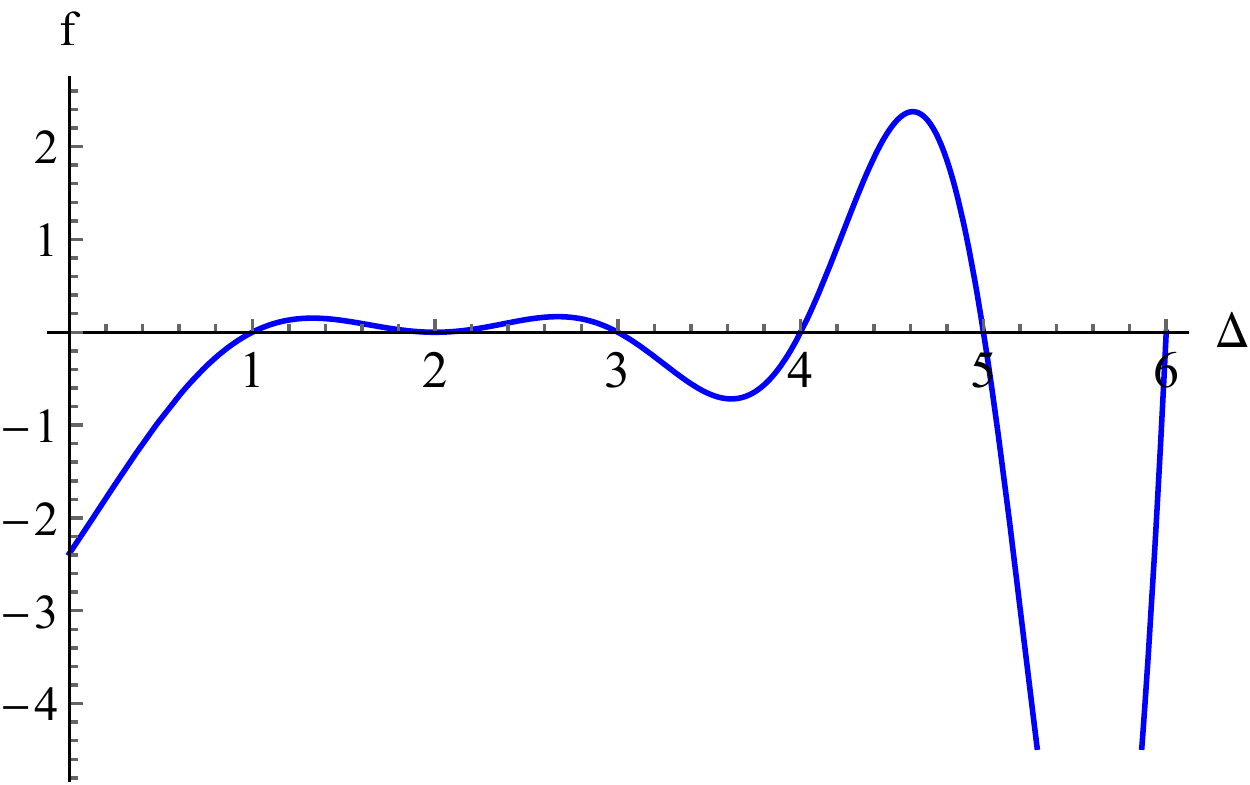}~~~
\includegraphics[width=0.48\linewidth]{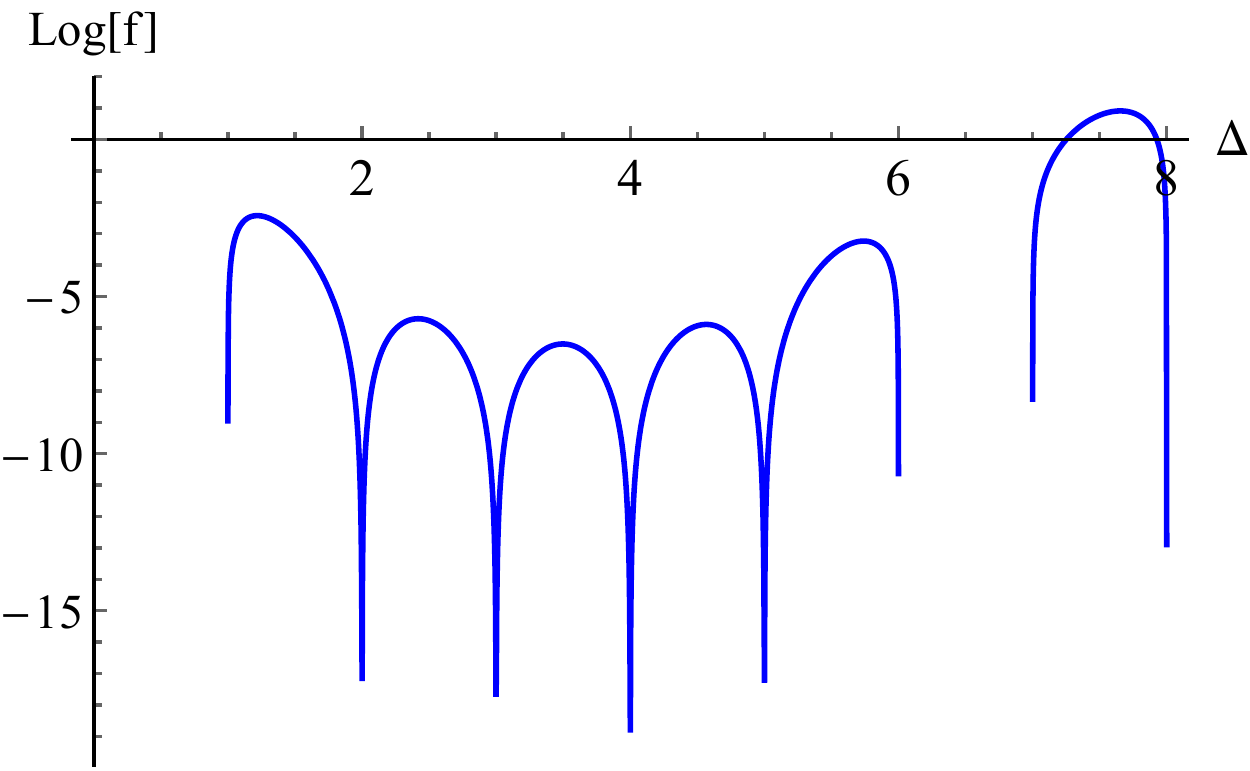}
\caption{Actions (denoted $f$) of the functionals $\alpha_N'$ on the t-channel conformal block with $N=1$ (left) and $N=4$ (right). Note in the right plot, the $y$-axis is $\log(f)$ which has no real value for negative $f$. } \label{FuncAction}
\end{figure}

\begin{figure}
\begin{center}
   \includegraphics[width=0.6\linewidth]{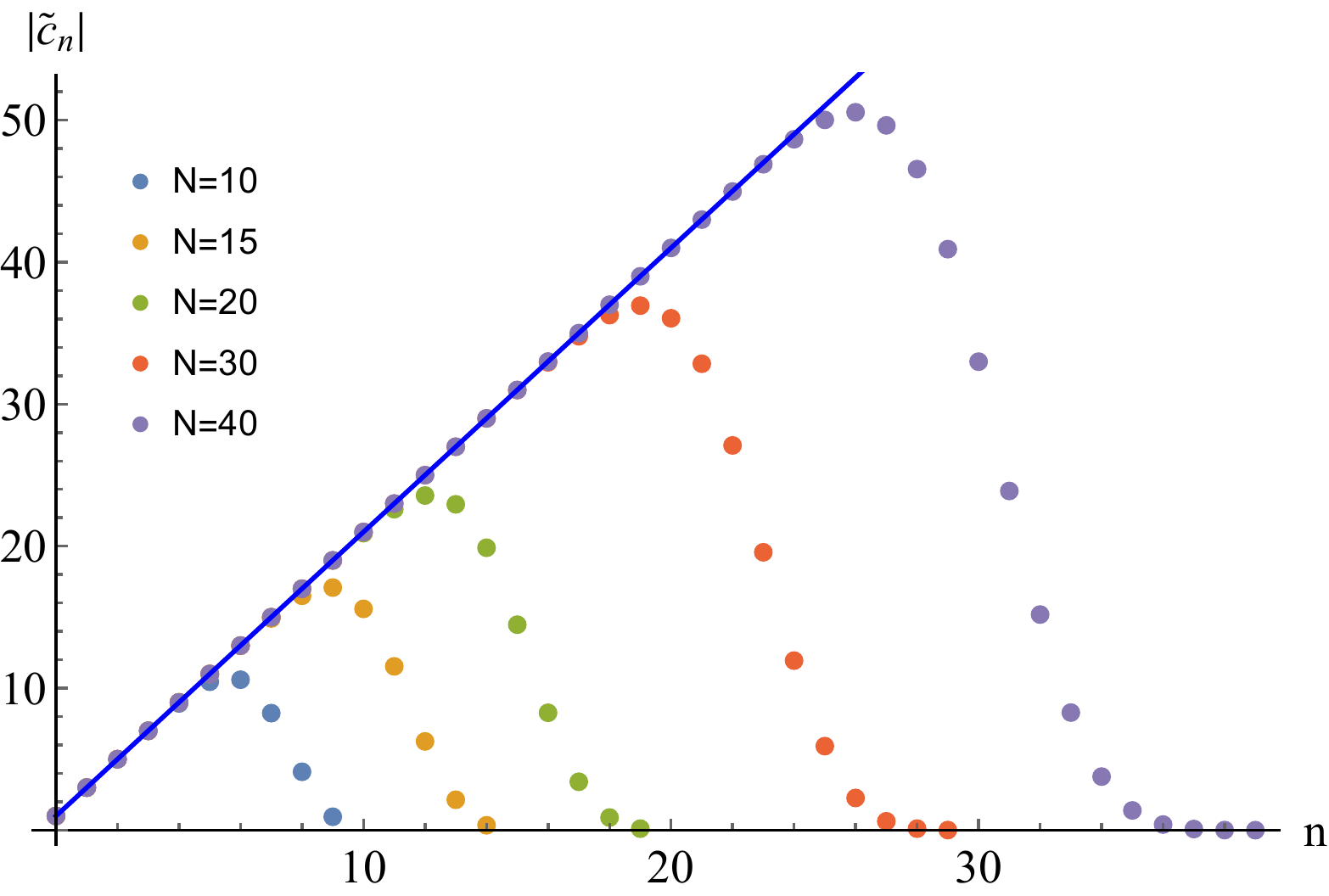}  
\end{center}
\caption{$|\Tilde{c}_n|$ solved from (\ref{eqc1}) with finite $N$'s. The straight line is $|\Tilde{c}_n|=2n+1$.} \label{cntilde}
\end{figure}

It is straightforward to solve above equations   for small $N$'s.
Taking $N=4$, the matrix $\mathcal{T}(n+1,i)$  is\footnote{The cautious readers may notice that the  matrix is symmetric up to certain numerical factors. It can be proved that the matrix $M_{m,n}\equiv \frac{\Gamma(m+1)^2}{\Gamma(2m+2)}\mathcal{T}(m+1,n)$ is indeed symmetric.}
{\footnotesize
\begin{align}
   \left(
\begin{array}{ccccc}
 \frac{\pi ^2}{6} & \frac{1}{6} (12-\pi ^2) & \frac{1}{6} (\pi ^2-9) & \frac{1}{18} (31-3 \pi ^2) & \frac{1}{72} (12 \pi ^2-115) \\
 12-\pi ^2 & 5 \pi ^2-48 & 129-13 \pi ^2 & \frac{1}{3} (75 \pi ^2-739) & \frac{1}{12} (4859-492 \pi ^2) \\
 5 (\pi ^2-9) & 5 (129-13 \pi ^2) & 5 (73 \pi ^2-720) & \frac{5}{3} (7492-759 \pi ^2) & \frac{5}{12} (7932 \pi ^2-78283) \\
 \frac{70}{9} (31-3 \pi ^2) & \frac{70}{9} (75 \pi ^2-739) & \frac{70}{9} (7492-759 \pi ^2) & \frac{70}{9} (4335 \pi ^2-42784) & \frac{35}{18} (679939-68892 \pi ^2) \\
 \frac{35}{4} (12 \pi ^2-115) & \frac{35}{4} (4859-492 \pi ^2) & \frac{35}{4} (7932 \pi ^2-78283) & \frac{35}{4} (679939-68892 \pi ^2) & 35 (99003 \pi ^2-977120) \\
\end{array}
\right), \nn
\end{align}}

\noindent{}which corresponds to the coefficients $\Tilde{c}_i$  
\begin{equation}
\Tilde{c}_i\approx(0.999753,\; -2.97534,\; 4.5689,\; -4.39935,\; 1.88092).
\end{equation}
It is  impressive that the first few elements are close to the limit $\Tilde{c}_n=(-1)^{-n}(2n+1)$ even for $N=4$. In Fig. \ref{cntilde} we show more solutions of $\Tilde{c}_n$ with larger $N$'s.\footnote{To solve $\Tilde{c}_n$ from (\ref{eqc1}) with large $N$, it is necessary to adopt high numerical precision, reminiscent of the numerical conformal bootstrap with SDPB \cite{Simmons-Duffin:2015qma, Landry:2019qug}.}
From these examples, the coefficients $|\Tilde{c}_n|$ with $n\leqslant N/2$ are close to the limit $|\Tilde{c}_n|= (2n+1)$, while for larger $n$'s, $|\Tilde{c}_n|$  deviates the straight line and decreases exponentially. Nevertheless, for all $n$'s the coefficients $\Tilde{c}_n$ have signs $\Tilde{c}_n\propto (-1)^n$. This suggests the coefficients $c_n\propto(-1)^n\Tilde{c}_n$ are all positive, therefore satisfying the positive condition (\ref{cd2}) up to $\Delta<N+1$.

\begin{figure}
  \begin{center}
      \includegraphics[width=0.75\linewidth]{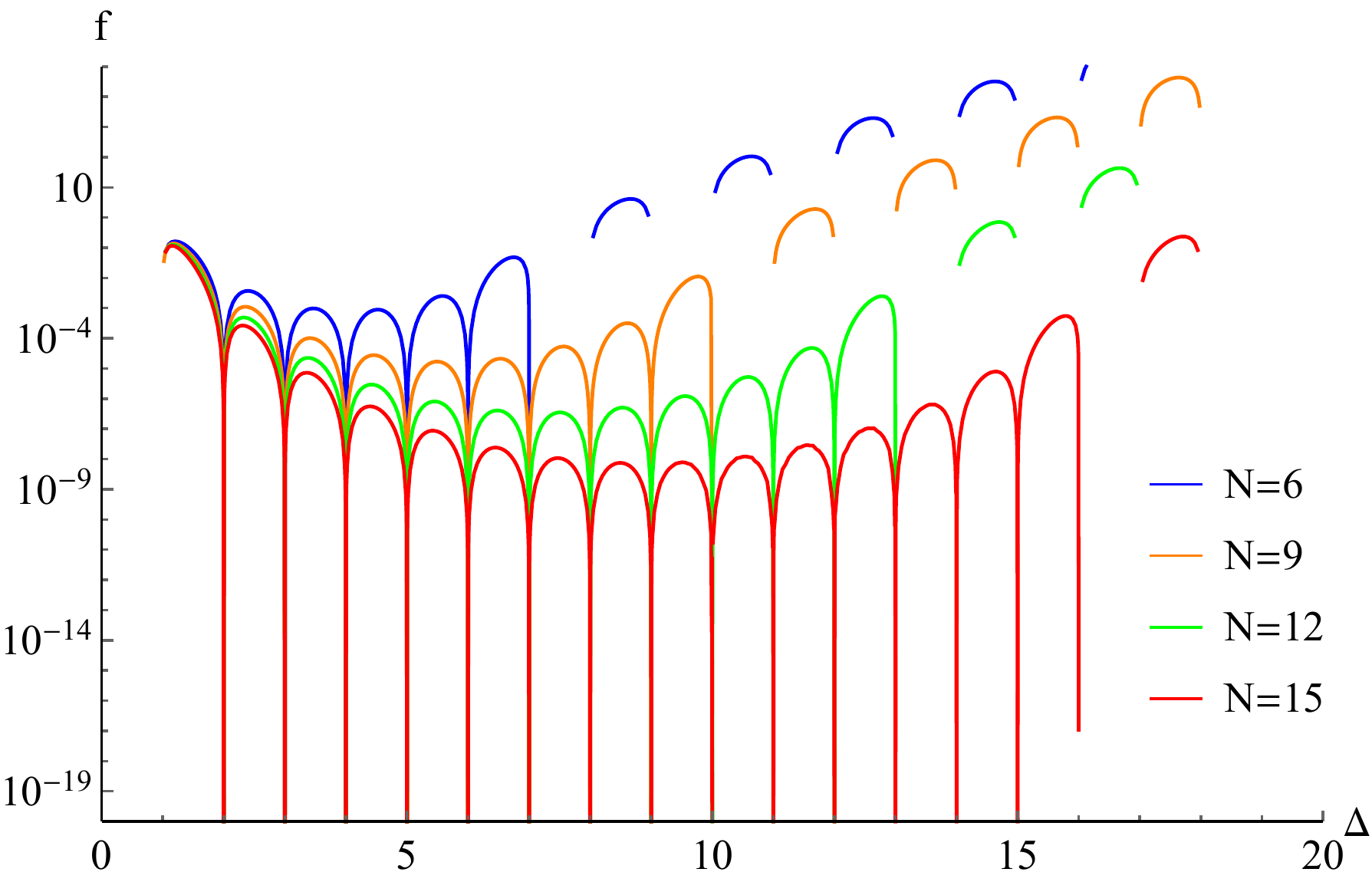}  
  \end{center}
\caption{Actions of the functionals $\alpha_{N}'$ on the t-channel conformal block. } \label{MoreFunc}
\end{figure}

\begin{figure}
\begin{center}
   \includegraphics[width=0.6\linewidth]{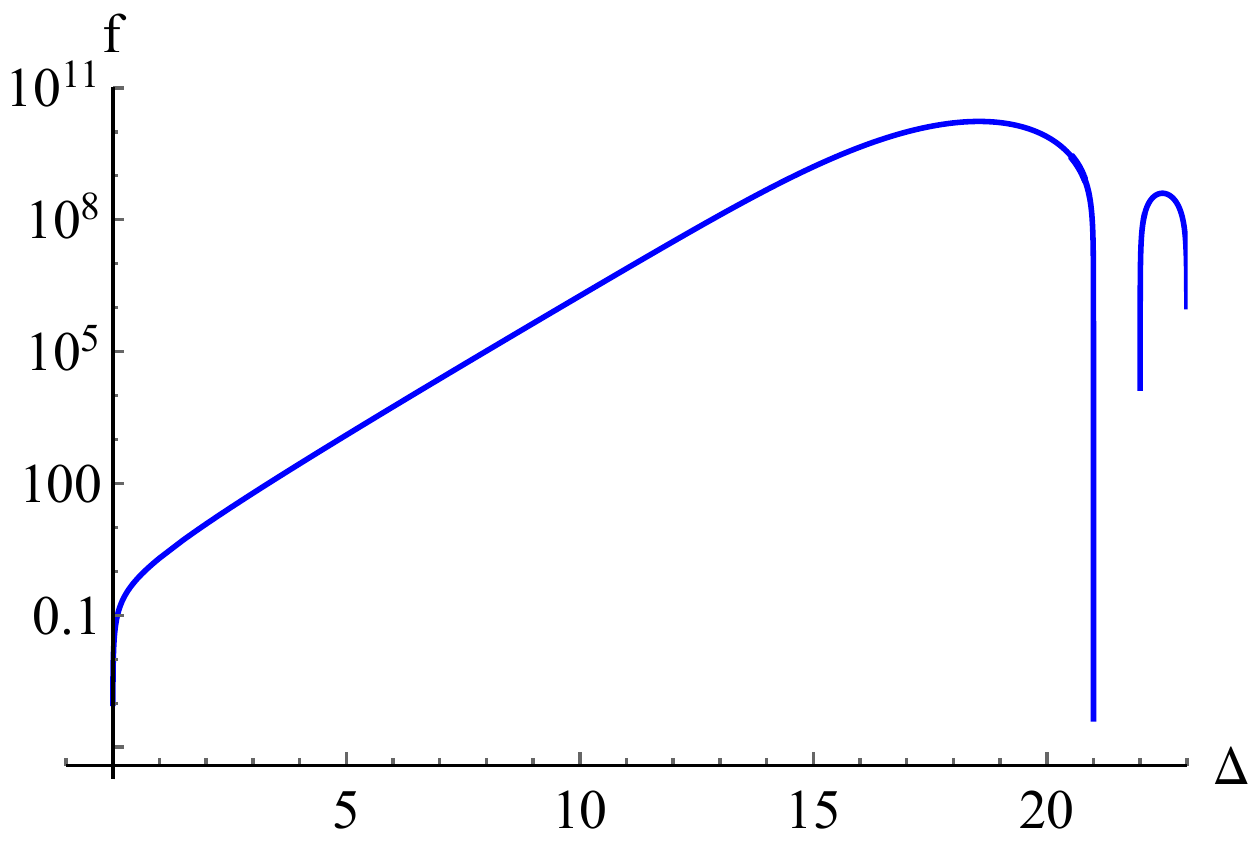}  
\end{center}
\caption{Action of the functional 
$\alpha_{20}'$ on the s-channel conformal block. } \label{SingletFunc}
\end{figure}

In Figs. \ref{MoreFunc} and \ref{SingletFunc} we show the actions (denoted $f$) of the functional $\alpha_N'$ on the t-channel and s-channel conformal blocks, respectively. The actions on the t-channel conformal block have a single zero at $\Delta=1$ and double zeros at $\Delta=n+1$ for integer $0<n<N$.
For $1<\Delta<N+1$, the $f$ decreases with larger $N$. This is consistent with our previous results that with $N=\infty$, the functional $\alpha^*$ becomes trivial: $f=0$  for $\Delta\geqslant2\Delta_\phi$. The action on the s-channel conformal block ($N=20$) shown in Fig. \ref{SingletFunc}  has a single zero at $\Delta=0$, corresponding to the unit operator, and is positive for $0<\Delta<N+1$. The fact $\left.f(n)\right.|_{n>20}=0, ~n\in \mathbb{N}$  is expected since there is no $\alpha_{n>20}^s$ in $\alpha_{N=20}'$.
The numerical solutions of the equations (\ref{eqc1}) do satisfy all the positive conditions (\ref{cd1}-\ref{cd5}) up to $\Delta<N+1$ for finite $N$.

\subsubsection{Positivity from total positivity}
Our goal is to generalize previous results to arbitrarily large $N\in \mathbb{N}^+$ with general $\Delta_\phi$. However, it requires highly nontrivial conditions to guarantee the strong positive constraints (\ref{cd1}-\ref{cd5}) for large $\Delta$.
The reason why our previous examples can satisfy the positive conditions for $\Delta\leqslant\Lambda_n$ is due to a  simple fact: in the action
\begin{equation}
    f(\Delta)\propto \sin(\pi(\Delta-2\Delta_\phi) \sum_{n=0}^N \Tilde{c}_n\mathcal{T}(\Delta,n),
\end{equation}
the sign of the function $\sum_{n=0}^N \Tilde{c}_n\mathcal{T}(\Delta,n)$ oscillates in phase with $\sin(\pi(\Delta-2\Delta_\phi)$ in the range $2\Delta_\phi<\Delta<2\Delta_\phi+N$. This is equivalent to the following properties of the function $\sum_{n=0}^N \Tilde{c}_n\mathcal{T}(\Delta,n)$ with general $N$:
\begin{itemize}
    \item The matrix $\left.\mathcal{T}(2\Delta_\phi+n,i)\right|_{0\leqslant n, i\leqslant N}$ is non-degenerate and invertible.
    \item All the zeros at $n=1,\dots, N$ are of order 1 or higher odd numbers.
    \item There are no other zeros besides $n=1,\dots, N$.
\end{itemize}
Any violations of above properties will necessarily break the positive conditions and invalid the functionals constructed from (\ref{eqc1}).  Surprisingly, above three properties are closely related to the total positivity of the $\SL$ conformal block for which we have studied in Section \ref{sec3}. 

Consider the equation (\ref{eqc1}) for general $\Delta_\phi$. The left part corresponds to 
\begin{align}
   g&(\Delta) \equiv \sum_{n=0}^N \Tilde{c}_n
\mathcal{T}(\Delta,n) = \nn\\    &\int_0^1dx\, (1-x)^{2\Delta_\phi-1}x^{-2\Delta_\phi}G_\Delta(x) \sum_{n=0}^N \Tilde{c}_n\, _3F_2(1,-n,4 \Delta_\phi +n-1;2 \Delta_\phi ,2 \Delta_ \phi ;1-x), \label{IntegralGD}
\end{align}
where
\begin{equation}
    \Tilde{c}_n= c_n\frac{(-1)^n  (2 \Delta_\phi )_n ^2}{n! (n+4 \Delta_\phi -1)_n}.
\end{equation}
In (\ref{IntegralGD}), the total positivity of the factor $(1-x)^{2\Delta_\phi-1}x^{-2\Delta_\phi}G_\Delta(x)$ follows the total positivity of the conformal block $G_\Delta(x)$  for $\Delta>\Delta_{\textrm{TP}}^*$. Therefore we have the {\it Variation Diminishing Property}  (\ref{VDP1}) that the sign changes of the functions $g(\Delta)$ and
${\Bar{\Theta}}(x)=\sum\Tilde{c}_n\;_3F_2$ in (\ref{IntegralGD}) should satisfy the relation
\begin{equation}
    S^+(g)\leqslant S^+({\Bar{\Theta}})\leqslant N, \label{signchanges}
\end{equation}
i.e., the number of sign changes of $g(\Delta)$ in $\Delta\in (2\Delta_\phi,\infty)$\footnote{In general the lower bound of $\Delta$ should be the critical value $\Delta_{\textrm{TP}}^*$ where $G_\Delta(x)$ loses its total positivity. Here we assume $2\Delta_\phi>\Delta_{\textrm{TP}}^*$ and the problem with smaller $\Delta_\phi$ will be studied later.} is not larger than the number of sign changes of $\Bar{\Theta}(x)$ in $x\in(0,1)$! The second inequality in (\ref{signchanges}) is due to the fact that $\Bar{\Theta}(x)$ is a polynomial of order $N$
\begin{equation}
    \Bar{\Theta}(x)=\sum_{n=0}^Na_n x^n, \label{Thetax}
\end{equation}
consequently including multiplicity, there are  at most  $N$ zeros of $ \Bar{\Theta}(x)$. Only the odd order zeros relate to sign changes of $ \Bar{\Theta}(x)$, therefore the polynomial $ \Bar{\Theta}(x)$ can have at most $N$ sign changes, corresponding to $N$ first order zeros. Moreover, in the extremal case $S^+(\Bar{\Theta})=N$,
according to the {\it Descartes' rule of signs},  the number of positive roots of $ \Bar{\Theta}(x)$ is at most the number of sign changes in the sequence of its coefficients $\{a_n\}$, therefore there has to be $N$ sign changes in  $\{a_n\}$, which requires $a_n\propto a_0 (-1)^n$.

The inequality (\ref{signchanges})
provides substantial restrictions on the solutions of the equation group (\ref{eqc1})!
The function $g(\Delta)$ has at most $N$ zeros including multiplicity. Therefore the $N$ zeros specified in the equations (\ref{eqc1}) at $\Delta=2\Delta_\phi+n,~ n=1,2,\dots,N$ are all  the zeros allowed by the inequality (\ref{signchanges}). Moreover, they are single zeros! With $S^+(g)= S^+({\Bar{\Theta}})$, the {\it Variation Diminishing Property} requires the function $\Bar{\Theta}(x)$ should have the same sign arrangement in $x\in(0,1)$ as $g(\Delta)$ in $\Delta\in(2\Delta_\phi,\infty)$, which is in the order $\{+,-,\cdots\}$ according to the equation (\ref{eqc1}). This suggests $a_0>0$ and consequently, $a_n\propto a_0(-1)^n \propto (-1)^n$. In particular, we have 
\begin{equation}
  c_N\propto (-1)^N\Tilde{c}_N\propto (-1)^N a_N>0.  \label{cN}
\end{equation}

For small $\Delta_\phi$, e.g. $\Delta_\phi=0.1$, the conformal block $G_\Delta(x)$ is not totally positive at $\Delta=2\Delta_\phi$, and the sign inequality (\ref{signchanges}) only works for $\Delta\in(\Delta_{\textrm{TP}}^*,\infty)$.
The numerical computations indicate  $\Delta_{\textrm{TP}}^*\approx 0.32315626<1$, which implies all the zeros  of $g(\Delta)$ specified in (\ref{eqc1}): $\Delta=2\Delta_\phi+n, 1\leqslant n\leqslant N$
are above  $\Delta_{\textrm{TP}}^*$ and the inequalities (\ref{signchanges}) can prove that these zeros are of first order and no other zeros in $(\Delta_{\textrm{TP}}^*,\infty)$. Besides, one  needs to show that there are no extra zeros of $g(\Delta)$ between $(2\Delta_\phi, \Delta_{\textrm{TP}}^*)$. We do not have a strict proof for this statement but have verified it by numerically solving (\ref{eqc1}) with small $\Delta_\phi$'s.

We also need to prove that there are indeed $N$ zeros in $g(\Delta)$, e.g., the equations (\ref{eqc1}) have non-trivial solutions. This is equivalent to the statement that the matrix $\left.\mathcal{T}(2\Delta_\phi+n,i)\right|_{0\leqslant n, i\leqslant N}$ is  invertible for any $N$, which can be proved within two steps based on the assumed total positivity of the function $f(\Delta,i)=(\Delta)_i^2/(2\Delta)_i$. Firstly it can be shown that the integral 
\begin{equation}
    \mathcal{P}(\Delta,k)=\int_0^1dx (1-x)^{2\Delta_\phi-1}x^{-2\Delta_\phi}G_\Delta(x) x^k
\end{equation}
is totally positive,
similar to (\ref{TPGaction}). Therefore the determinant of its sub-matrices are always nonzero. The determinant $||\mathcal{T}(\Delta,k)||_{N}$ is related to $||\mathcal{P}(\Delta,k)||_N$ through a 
non-degenerate basis transformation $\{x^k\}\rightarrow \{\Theta_k\}$ and is also nonzero. 

Therefore based on the total positivity of the $\SL$ conformal block, 
previous three questions on the equation group (\ref{eqc1}) can be nicely addressed. It suggests that for general $N$, the equations (\ref{eqc1}) always have nontrivial solutions, and the related function $g(\Delta)$
only has single zeros at $\Delta=2\Delta_\phi+n, ~n=1,...,N$. This guarantees the positive conditions on the t-channel conformal block (\ref{cd3}-\ref{cd5}) for $\Delta\leqslant 2\Delta_\phi+N$ with arbitrary positive integer $N$.

An interesting question is whether the matrix $\mathcal{T}(\Delta, n)$ is also totally positive. If true, then it can prove that the coefficients $c_n$ solved from (\ref{eqc1}) are always positive. 
The {\it Variation Diminishing Property } tells us the signs of $a_n$ in (\ref{Thetax}) with basis $\{x^n\}$, which can  be used to verify the positive sign of $c_N$ (\ref{cN}). 
In future studies, it would be important to provide quantitative estimations of $\Tilde{c}_n$'s for general $N$ and  explain the non-monotonic shapes in Fig. \ref{cntilde}.

~

Here we summarize the properties of the analytical functional $\alpha'_N=\sum_{n=0}^N c_n\alpha_n^s$
constructed through the equations (\ref{eqc1}): for a given positive integer $N$, the functional $\alpha'_N$ can produce the spectrum consistent with the numerical bootstrap results up to $\Delta\leqslant 2\Delta_\phi+N$. It gives the unit operator in the $O(N)$ singlet sector and  double trace operators in the $O(N)$ $T$ sector below $2\Delta_\phi+N$. The actions of the functional satisfy the positive conditions in both singlet and 
$T$ sectors for $\Delta<2\Delta_\phi+N$.
Based on the total positivity of the $\SL$ conformal block, such functionals exist for any finite $N$. The large $N$ limit of the functional 
$\lim\limits_{N\rightarrow \infty}\alpha'_N= \alpha^*$ 
is trivial, which produces zero action on the t-channel conformal block. In contrast, 
the truly nontrivial point here is the way how the series of functionals $\{\alpha'_N\}$ approach the limit $\alpha^*$: for any given large cutoff $\Lambda_N$, one can construct $\alpha_N'$ so that its action satisfies the required positive conditions for any $\Delta< \Lambda_N$. This explains, for the Regge superbounded correlators, the numerical bootstrap bound of the crossing equation (\ref{simpceq}), or the bound with  $\Delta_\phi<\Delta_c/2$ in Fig. \ref{BD1D}.

\subsection{Analytical functionals for general conformal correlators}
The functional constructed in the last section scales as $O(|z|^{-1})$ in the Regge limit and only works for the superbounded conformal correlators, e.g. (\ref{GFF1}-\ref{GFF3}) with $\lambda=0$. For more general correlators, such as (\ref{GFF1}-\ref{GFF3}) with $\lambda\neq0$, one can construct functionals  using the subtracted basis  $\Bar{\alpha}_n^{s}$ (\ref{alphabar}):
\begin{equation}
    \Bar{\alpha}_N'=\sum_{n=1}^Nc_n \Bar{\alpha}_n^s.
\end{equation}
Above functional should satisfy the same positive conditions (\ref{cd1}-\ref{cd5}). Following the same reasons for (\ref{eqc1}) we can get a similar equation group
\begin{equation}
    \sum_{i=1}^N\Bar{\mathcal{T}}(2\Delta_\phi+n,i)\cdot \tilde{c}_i =\delta_{n0}, ~~\textrm{for} ~n=0,1,\dots,N-1, \label{eqc2}
\end{equation}
in which
\begin{equation}
    \Bar{\mathcal{T}}(\Delta,i)=\mathcal{T}(\Delta,i)-\frac{(-1)^i(2\Delta_\phi)_i^2}{i!(4\Delta_\phi+i-1)_i}\mathcal{T}(\Delta,0)
\end{equation}
 is invertible. 
 Solutions to (\ref{eqc2}) are related to the function $\bar{g}$
\begin{align}
   &\bar{g}(\Delta) \equiv \sum_{n=1}^N
\bar{\mathcal{T}}(\Delta,n) \Tilde{c}_n=  \label{IntegralGD1} \\    &\int_0^1dx (1-x)^{2\Delta_\phi-1}x^{-2\Delta_\phi}G_\Delta(x) \sum_{n=1}^N \Tilde{c}_n\left(\, _3F_2(1,-n,4 \Delta_\phi +n-1;2 \Delta_\phi ,2 \Delta_ \phi ;1-x)-1\right). \nn
\end{align}
Again we want to bound the number of zeros of the function  $\Bar{g}(\Delta)$ by the number of sign changes in the sequence of the polynomial $\Tilde{\Theta}(x)=\sum \Tilde{c}_n (\,_3F_2-1)$. However, the function $\Bar{g}(\Delta)$ is expected to have $N-1$ zeros at $\Delta=2\Delta_\phi+n$ with $n=1,...,N-1$, while $\Tilde{\Theta}(x)$ remains an order $N$ polynomial of $x$, which in principle could have $N$ single zeros and sign changes, indicating the function $\Bar{g}(\Delta)$ could have, including multiplicity, an extra zero besides the $N-1$ single zeros specified in (\ref{eqc2}). Solution to this puzzle is that the lowest term of $\Tilde{\Theta}(x)$, when expanded as an order $N$ polynomial of $(1-x)$, is linear in  $(1-x)$, while the constant term has been canceled in the subtraction (\ref{alphabar}) for better Regge behavior. Therefore this linear term can be factorized and $\Bar{g}(\Delta)$ becomes
\begin{equation}
    \bar{g}(\Delta)=\int_0^1dx (1-x)^{2\Delta_\phi}x^{-2\Delta_\phi}G_\Delta(x) \sum_{n=0}^{N-1} \lambda_n x^n.
\end{equation}
The function $(1-x)^{2\Delta_\phi}x^{-2\Delta_\phi}G_\Delta(x)$ remains totally positive while the order of the polynomial  
$\Tilde{\Theta}(x)$ is reduced to $N-1$, which can have at most $N-1$ zeros and sign changes. Therefore according to 
the {\it Variation Diminishing Property } (\ref{VDP1}), the function $\Bar{g}(\Delta)$ can have at most $N-1$ zeros. This confirms the function $\Bar{g}(\Delta)$ has no other zeros besides $2\Delta_\phi+n$, $n=1,...,N-1$. Moreover, they are single zeros. 
It can be verified numerically that the coefficients $c_n$ solved from (\ref{eqc2}) are all positive, corresponding to positive actions of $\bar{\alpha}_N'$ on the s-channel conformal for $\Delta\in(0,2\Delta_\phi+N)$, similar to Fig. \ref{SingletFunc}.

The conclusion is that the subtraction (\ref{alphabar}) is consistent with {\it Variation Diminishing Property} and the  functionals $\bar{\alpha}_N'$  for the general conformal correlators have similar positive properties as the functionals  $\alpha_N'$  for the Regge superbounded correlators.

\section{Conclusion and Outlook}\label{sec5}
We have studied the 1D $O(N)$ vector bootstrap in the large $N$ limit. 
    We  obtained a remarkably simple  bootstrap equation with bootstrap bound saturated by the generalized free field theory. The most interesting part of this work is the construction of analytical functionals for the large $N$ bootstrap. We proposed an approach to construct a series of bootstrap functionals $\{\alpha_M'\}$ whose actions on the crossing equations can satisfy the bootstrap positive conditions for $\Delta\leqslant \Lambda_M$. A surprising fact is that although the large $M$ limit of the functionals $\alpha_M'$ becomes trivial, the functionals $\{\alpha_M'\}$ can approach the limit in  a particular way so that the bootstrap positive conditions can be fulfilled at arbitrarily high $\Lambda_M$, thus providing an analytical explanation for the bootstrap bound. We found the total positivity of the $\SL$ conformal block relates to a sophisticated mathematical structure and plays a substantial role to construct  analytical functionals. This work provides a concrete example to illustrate the mathematical  structure in conformal bootstrap and the intriguing connections between mathematics and quantum field theories.

We believe this work opens the door towards more systematical studies for many fascinating problems in quantum field theories and their connections to mathematics. Part of these problems are explained below.
\begin{itemize}
    \item  The most fundamental question is the total positivity of the $\SL$ conformal block $G_\Delta(z)$, which provides the key ingredient in bootstrap studies. We have proved the $\SL$ conformal block is totally positive with large $\Delta$ and showed the total positivity is violated below a threshold value $\Delta_{\textrm{TP}}^*\approx 0.32315626$. We have provided numerical evidence indicating this estimation could be optimal but a strict proof is not available yet. Moreover, we have observed that total positivity of the conformal block relates to a special mathematical structure which can naturally generate a huge hierarchy in the parameter space. It would be exciting to improve our understanding of this mathematical structure and its  applications in quantum field theories.
    
    \item  In this work we have constructed the analytical functional for the first part of the 1D large $N$ bootstrap bound before the kink in Fig. \ref{BD1D}, which is saturated by the generalized free field theory and the bootstrap equations are reduced to a simple form (\ref{simpceq}).  It is  tempting to know the theories saturating the second part of the bootstrap bound and construct the analytical functionals. Furthermore, the bootstrap  bound almost disappears after $\Delta_\phi>0.75$ in Fig. \ref{BD1D}. Similar phenomenon also appears in higher dimensions, see  Fig. \ref{BD23D}. It would be interesting to know the reasons which dissolve  the bootstrap restrictions.
    
    \item  Conformal field theories with large $N$ limits have close relation to the quantum field theories in the AdS spacetime. Constraints on the CFT side can lead to nontrivial restrictions on the theories in AdS, see e.g. \cite{Paulos:2016fap, Antunes:2021abs, Caron-Huot:2021enk, Cordova:2022pbl, Knop:2022viy}. It would be interesting to explore the constraints of the analytical functional constructed in this work on the S-matrices in AdS$_2$. In particular, how does the total positivity affect the scattering process in AdS$_2$? Do the AdS analogs of the conformal blocks, the Witten diagrams also satisfy total positivity?
The role of total positivity in the 4D amplitude in flat spacetime has been extensively studied recently \cite{Arkani-Hamed:2012zlh,Arkani-Hamed:2013jha,Arkani-Hamed:2020blm,Herrmann:2022nkh}. Our results on the 1D CFTs suggest that the AdS$_2$ could provide another interesting and technically tractable laboratory to explore the role of (total) positivity in quantum field theories. 
We hope to report the applications of analytical functionals and total positivity on AdS physics in another work.

    \item Total positivity is   powerful  to analyze positivity of analytical functionals. In our construction, the positivity of bootstrap functionals can be established based on the total positivity of the $\SL$ conformal block while without solving the equation groups (\ref{eqc1},\ref{eqc2}) explicitly. Nevertheless, it would be interesting to know more concrete information on the  analytical functionals $\{\alpha_M'\}$, such as the curves of the coefficients $|\Tilde{c}|_n$ shown in Fig. \ref{cntilde}. One may wonder if the equation groups (\ref{eqc1}, \ref{eqc2}) are easier to solve in Mellin space \cite{Penedones:2010ue, Fitzpatrick:2011ia,Paulos:2011ie}. 
    
    \item The 1D large $N$ vector bootstrap provides
insights to study higher dimensional $O(N)$ vector bootstrap. There are solid evidence for close relations between the two problems. Firstly their bootstrap bounds have similar patterns, as shown in Figs. \ref{BD1D} and \ref{BD23D}. Moreover, for the bootstrap bounds saturated by the generalized free field theories, the $O(N)$ vector bootstrap equations degenerate to  similar forms in 1D and higher dimensions. The functional basis dual to higher dimensional generalized free field spectrum has been constructed in \cite{Mazac:2019shk} and their relation to dispersion relation has been studied in \cite{Caron-Huot:2020adz}, see also \cite{Bissi:2019kkx, Paulos:2019gtx}. However, a crucial question is how to organize the functional basis in order to satisfy the   positive conditions. The method developed in this work can be useful to construct analytical functionals with suitable positive properties in higher dimensions. We leave this problem for future work \cite{ZL23}.

    \item  A more challenging problem  along this direction is to construct the analytical functionals for the $O(N)$ vector bootstrap bounds with large but finite N. This was one of the motivations for the author to start this work.
In this case we need to go back to the whole $O(N)$ vector crossing equations (\ref{fdcr0},\ref{fdcr}) and take   the $1/N$ terms into account.
These $1/N$ terms and the crossing equation (\ref{fdcr0}) will necessarily introduce new ingredients responsible for the $1/N$ interactions in the underlying theories. The related analytical functionals could provide a new nonperturbative frame to study CFTs with large $N$ limits, including the 3D critical $O(N)$ vector models and  the conformal gauge theories in general dimensions. 

    \item The series of analytical functionals $\{\alpha_N'\}$ constructed in this work are
sensitive to the large $\Delta$ spectrum. Associated with total positivity, they can be employed to detect non-unitarity in the large $\Delta$ region, which relates to the high energy dynamics in AdS. We hope more systematical studies of the large $N$ analytical functionals can provide solid conclusions for some widely interested questions on the large $\Delta$ spectrum of large $N$ unitary CFTs. 

\end{itemize}

\section*{Acknowledgements}
The author would like to thank Nima Arkani-Hamed, Greg Blekherman, Miguel Paulos and David Poland for discussions. 
The author is grateful to David Poland for the valuable support. The author thanks the organizers of the conferences ``Bootstrapping Nature: Non-perturbative Approaches to Critical Phenomena" at Galileo Galilei Institute, ``Positivity" at Princeton Center for Theoretical Science and Simons Collaboration on the Nonperturbative Bootstrap Annual Meeting for creating
stimulating environments.
This research was supported by Shing-Tung Yau Center and Physics Department at Southeast University, 
Simons Foundation grant 488651 (Simons Collaboration on the Nonperturbative Bootstrap) and DOE grant DE-SC0017660.  The bootstrap computations were carried out on the Yale Grace computing cluster, supported by the facilities and staff of the Yale University Faculty of Sciences High Performance Computing Center.

\appendix
\section{Examples of the totally positive functions} \label{TPexamples}
In this appendix we show some classical examples of the totally positive functions. Some of the results in this part have been applied in our study of the total positivity of the Gauss hypergeometric functions $_2F_1(\Delta,\Delta,2\Delta,z)$ and  $\SL$ conformal block functions
$G_\Delta(z)$.

\subsection{Example 1: $f(\Delta,x)=x^\Delta$} \label{TPEx1}
 
The determinant formula (\ref{TPdf}) of the function $f(\Delta,x)=x^\Delta$ is given by
\bea 
||f(\Delta,x)||_m\equiv f\left(
\begin{array}{ccc}
    \Delta_1, & ... & \Delta_m \\
    x_1, & ... & x_m
\end{array}
\right)=\det \left[\begin{array}{ccc}
   x_1^{\Delta_1}  & ... & x_1^{\Delta_m}  \\
   \vdots & & \vdots \\
   x_m^{\Delta_1}  &...& x_m^{\Delta_m} 
\end{array}
\right].  
\eea 
Taking $\Delta_i=i-1$, above determinant goes back to the Vandermonde determinant, which is given by 
\beq
||f(\Delta,x)||_m=\prod_{i>j}(x_i-x_j)
\eeq
and is positive for the ordered variables $0<x_1<\dots<x_m$. Then to prove the total positivity of the function $f(\Delta,x)$, one only needs to show that its determinant can never be zero, which can be done by induction \cite{Gantmacher1961OscillationMA,Arkani-Hamed:2018ign}.

The statement $||f(\Delta,x)||\neq0$ is equivalent to the claim that for a given set of $c_i\in \mathbb{R}$, the equation
\begin{equation}
    h_m(x)=\sum_{i=1}^m c_i x^{\Delta_i}
\end{equation}
cannot have $m$ solutions in the region $x>0$. For $n=1$, $h_1(x)=c_1 x^{\Delta_1}$ and there is no positive solution for $h$. Assume above statement is true for $h_i(x)$ with $i<n$. If $h_n(x)$ has $n$ positive solutions, then according to Rolle's theorem, the following function
\begin{equation}
   (x^{-\Delta_1}h_n(x))'=\sum_{i=2}^{n}(\Delta_i-\Delta_1)c_i \, x^{\Delta_i-\Delta_1} \sim h_{n-1}(x) 
\end{equation}
can have $n-1$ positive zeros, which is inconsistency with our previous induction assumption that $h_{n-1}(x)$ cannot have $n-1$ positive solutions. Therefore the function $h_n(x)$ should have positive solutions less than $n$. This completes the proof that the determinant $||f(\Delta,x)||_m$ can never be zero.  

From the total positivity of the function $x^\Delta$, one can  show a family of totally positive functions. For instances, the function $e^{xy}=(e^x)^y$ is also totally positive.

\subsection{Example 2: $f(x,y)=\frac{1}{x+y}$} \label{TPEx2}
The determinant formula (\ref{CBf})
for the function $f(x,y)$ is
\bea 
||f(\Delta,x)||_m\equiv f\left(
\begin{array}{ccc}
    x_1, & ... & x_m \\
    y_1, & ... & y_m
\end{array}
\right)=\det \left[\begin{array}{ccc}
   \frac{1}{x_1+y_1}  & ... & \frac{1}{x_1+y_m}  \\
   \vdots & & \vdots \\
   \frac{1}{x_m+y_1}  &...& \frac{1}{x_m+y_m} 
\end{array}
\right].  
\eea 
Above determinant can be solved in a compact form, i.e., the Cauchy formula
\begin{equation}
    ||f(\Delta,x)||_m=\frac{\prod\limits_{i>k}(x_i-x_k)\prod\limits_{i>k}(y_i-y_k)}{\prod\limits_{i,k=1}^m(x_i+y_k)},
\end{equation}
which is positive for the ordered variables $x_1<\dots<x_m,~y_1<\dots<y_m$.

The total positivity of $f(x,y)$ can be alternatively proved using the basic composition formula (\ref{BCf}). The function can be rewritten as
\begin{equation}
    \frac{1}{x+y}=\int_0^\infty e^{-(x+y)t}dt=\int_0^1 u^x u^y d(\log(u)).
\end{equation}
Due to the basic composition formula, the total positivity of above integral follows the total positivity of the function $u^x$.

\bibliographystyle{utphys.bst}
\bibliography{LargeN1D}
\end{document}